\documentclass[12pt,aps,preprint,pre,amsmath,amssymb,showpacs,floatfix]{revtex4}
\usepackage{epsfig}
\newcommand{\Op}[1]{{\boldsymbol{\mathrm{\hat{#1}}}}}

\begin{document}

\title{The Quantum Four Stroke Heat Engine: 
Thermodynamic Observables in a  Model with Intrinsic Friction }

\author{Tova Feldmann and Ronnie Kosloff}
\email{ronnie@fh.huji.ac.il}
\affiliation{
Department of Physical Chemistry 
the Hebrew University, Jerusalem 91904, Israel\\
}


\begin{abstract}
The fundamentals of a quantum heat engine are derived from first principles.
The study is based on the equation of motion of a minimum set of operators
which is then used to define the state of the system. 
The relation between the quantum framework and thermodynamical observables is examined.
A four stroke heat engine model with a coupled two-level-system as a working fluid
is used to explore the fundamental relations. 
In the model used, the internal Hamiltonian does not commute with the external control 
field which defines the two adiabatic branches. Heat is transferred to the 
working fluid by coupling  to hot and cold reservoirs under constant field values.
Explicit quantum equation of motion for 
the relevant observables are derived on all branches. 
The dynamics on the heat transfer constant field branches is solved in closed form.
On the adiabats, a general numerical solution is used and compared with a particular
analytic solution. These solutions are combined to construct the cycle of operation.
The engine is then analyzed in terms of frequency-entropy and entropy-temperature graphs.
The irreversible nature of the engine is the result of finite heat transfer rates
and friction-like behavior due to noncommutability of the internal and external Hamiltonian.
\end{abstract}
\maketitle

\section{Introduction}
\label{sec:introduction}

Analysis of heat engine models has been  a major part of thermodynamic development.
For example Carnot's engine preceded the concepts of energy and entropy \cite{carnot1824}.
Szilard and Brillouin constructed a model engine which enabled them to resolve 
the paradox raised by Maxwell's demon \cite{szilard29,brillouin}. The subsequent insight enabled 
the unification of negative entropy with information. 
In the same  tradition, the present paper studies  a heat engine 
model with a quantum working fluid for
the purpose of tracing the microscopic origin of friction.
The function of a quantum heat engine as well as its classical counterpart is to transform 
heat into useful work. In such engines, the work is extracted by an external field
exploiting the spontaneous flow of heat from a hot to a cold reservoir. 
The present model performs this task by a four stroke cycle of operation.
All four branches of the cycle can be described by quantum equations of motion.
The thermodynamical consequences can therefore be derived from first principles.

The present paper lays the foundation for a comprehensive analysis of a
discrete  model of a quantum heat engine. A brief outline which has been published 
emphasized the engines optimal performance characteristics \cite{kosloff01}. 
It was shown that the engines power output vs. cycle time mimics very closely a classical heat engine subject
to friction. The source of the apparent friction was traced back to a quantum phenomena:
the noncommutability of the external control field Hamiltonian and the 
internal Hamiltonian of the working medium. 

The fundamental issue involved require a detailed and careful study.
The approach followed  is to derive the thermodynamical concepts from quantum principles.
The connecting bridges are the quantum thermodynamical observables. 
Following the tradition of Gibbs
a minimum set of observables is sought which are sufficient to characterize 
the performance of the engine. When the working fluid is in thermal equilibrium, 
the energy observable is sufficient to completely describe the state
of the system and therefore all other observables. During the cycle of operation the 
working fluid is  in a non-equilibrium state. In frictionless engines,
where the internal Hamiltonian commutes with the external control field, 
the energy observable is still sufficient to characterize the engine's cycle 
\cite{feldmann96,feldmann00}.
In the general case additional variables have to be added. For example in
the current model, a set of three quantum thermodynamic observables is sufficient 
to characterize the performance. With only two additional variables the 
state of the working fluid can be characterized also . A knowledge of the state is necessary
in order to  evaluate the entropy and the dynamical temperature. These 
variables are crucial in establishing a thermodynamic perspective.

The current investigation is in line with previous studies of quantum heat engines 
\cite{kosloff01,feldmann96,feldmann00,lloyd,gustic,k24,geva0,geva1,geva2,chen02,bender00,bender02,gordon01,jose,geva02}. 
All the studies of first principle quantum 
models  have conformed to the laws of thermodynamics. 
These models have been either continuous resembling
turbines \cite{geva2,gordon01,jose},
or discrete as in the present model \cite{kosloff01,feldmann96,geva0,geva1,geva2}.
Surprisingly the performance characteristics of the models were in 
close resemblance to their realistic counterparts. 
Real heat engines operate far from the  reversible conditions,
where the maximum power is restricted due to finite heat transfer \cite{curzon75},
internal friction and heat leaks \cite{benshaul79,salamon77,salamon80,gordon91,andresen83,bejan96,gordon98}. 
Analysis of the quantum models of heat engines,
based on a first principle dynamical theory, enable to pinpoint 
the fundamental origins of finite heat transfer,
internal friction and heat leaks. 

Studies of quantum  continuous heat engine models have revealed most
of the known characteristics of real engines. In accordance with finite 
time thermodynamics the power always exhibits a definite maximum \cite{salamon77}, and 
the performance has been limited by heat leaks \cite{jose}.
Finally indications of restrictions due to  friction like phenomena have been indicated \cite{geva2}.
The difficulty with the analysis is that it is very hard to separate the individual
contributions in the case of a continuous operating engine.

To facilitate the interpretation a four stroke discrete engine has been chosen for analysis.
The cycle of operation is controlled by the  segments of time that the engine
is in contact with a hot and cold bath and by the time interval required to vary the
external field. To simplify the analysis the time segments where the working fluid 
is in contact with the heat baths are carried out at constant external field.
Such a cycle of operation resembles the Otto cycle which is composed
of two {\em isochores} where heat is transfered and two {\em adiabats} where work is done.
This simplification allows to obtain the values of the thermodynamical 
observables during the cycle of operation from first principles in closed form.

\section{Quantum Thermodynamical observables and their dynamics}
\label{sec:formal}

The quantum thermodynamical observables constitute a set of variables which
are sufficient to completely describe the heat engine performance characteristics
as well as the entropy and temperature changes of its working medium.
The analysis of the performance requires a quantum 
dynamical description of the changes in the thermodynamical observables
during the engine's cycle of operation. 
The thermodynamical observables are  associated 
with the expectation values of operators of the working medium. 
Using the formalism of von Neumann, an expectation of an observable 
$\langle {\bf \hat A} \rangle $
is defined by the scalar product between the operator ${\bf \hat A}$
representing the observable and the density operator ${{\boldsymbol{ \mathrm {\hat \rho}}}}$ representing
the state of the working medium:
\begin{equation} 
\langle {\bf \hat A} \rangle ~=~ \left({\bf \hat A}\cdot {{{\boldsymbol{ \mathrm {  \rho}}}}} \right)
~=~ Tr \{ {\bf \hat A}^{\dag} ~ {{{\boldsymbol{ \mathrm {  \rho}}}}} \}~~~.
\label{eq:sclar}
\end{equation}

The dynamics of the working medium is subject to external change of 
variables as well as heat transport from the hot and cold reservoirs.
The dynamics is then described within the formulation
of quantum open systems \cite{lindblad76,alicki87}, where 
the dynamics is generated by the Liouville super operator ${\cal L}$
either as an equation of motion for the state ${{{\boldsymbol{ \mathrm {  \rho}}}}}$ (Schr\"odinger picture):
\begin{equation}
\dot { {{{\boldsymbol{ \mathrm {  \rho}}}}}}~~=~~ {\cal L}( {{{\boldsymbol{ \mathrm {  \rho}}}}})~~~,
\label{eq:liouville}
\end{equation}
or as an equation of motion for the operator (Heisenberg picture):
\begin{equation}
\dot {\bf \hat A}~~=~~ {\cal L}^* ( {\bf \hat A}) 
~+~ \frac{\partial {\bf \hat A}}{\partial t}~~~.
\label{eq:heisenberg}
\end{equation} 
The second part of the r.h.s. appears since the operator ${\bf \hat A}$
can be explicitly time dependent.

Significant simplification is obtained \cite{lendi01} when:
\begin{itemize}
\item{a) The operators of interest form an orthogonal set 
${\bf \hat B_i}$ i.e. 
\begin{equation}
\left({\bf \hat B_i}\cdot {\bf \hat B_j}\right)= \delta_{\bf ij}~~,
\label{eq:ortho}
\end{equation}
where   
${\bf \hat B_0}={\bf \hat I}$ is the identity operator.}
\item{b) The set is closed to the operation of ${\cal L}^*$.
\begin{equation}
\dot {\bf \hat B_i}~=~{\cal L}^*({\bf \hat B_i})=\sum_{\bf j} l_{\bf j}^{\bf i} {\bf \hat B_j}~~~,
\label{eq:lio24}
\end{equation}
where $l_{\bf i}^{\bf j}$ are scalar coefficients composing the matrix $\tilde L$. }
\item{c) The equilibrium density operator is a linear combination of the set:
\begin{equation}
 {{{{\boldsymbol{ \mathrm {  \rho}}}}}}^{eq}~=~\frac{1}{N} {\bf \hat I} + \sum_{\bf k} b_{\bf k}^{eq} {\bf \hat B_k}~~~,
\label{eq:lio25}
\end{equation}
where  $N$ is the dimension of the Hilbert space and
$b_{\bf k}^{eq}$ are the equilibrium expectation values of the
the operators, $\langle {\bf \hat B_k}^{eq} \rangle $      .}
\end{itemize}
The operator property of Eq. (\ref{eq:lio24}) allows a direct solution 
to the Heisenberg equation of motion (\ref{eq:heisenberg})
by diagonalizing the $\tilde L$ matrix, relating observables
$\langle {\bf \hat B_k} \rangle $  at time $t$ to observables at time $t+\Delta t$ that is
$\vec {\bf b}(t+\Delta t) = {\cal U}(\Delta t) \vec {\bf b}(t)$ where ${\cal U}=e^{\tilde L \Delta t}$
and $\vec {\bf b}$ is a vector composed from the expectation values of ${\bf \hat B_k} $
( for an example Cf. (\ref{PropagTNew})~).

The time dependent expectation values $\vec {\bf b}(t)$ and Eq. (\ref{eq:lio25}) 
can be employed to reconstruct the density operator:
\begin{equation}
{{{{\boldsymbol{ \mathrm {  \rho}}}}} }_R~~=~~\frac{1}{N} {\bf \hat I} + \sum_{\bf k} b_{\bf k} {\bf \hat B_k}~~~,
\label{eq:dens} 
\end{equation}
where the expansion coefficients become $b_{\bf k} =\langle {\bf \hat B_k} \rangle $.
Although the set ${\bf \hat B_k}$ is not necessarily complete, equation
(\ref{eq:dens}) will still be used as a reconstructing method for the density
operator. This reconstructed state ${{{{\boldsymbol{ \mathrm {  \rho}}}}} }_R$ reproduces all observations
which are constructed from linear combinations of the set of operators 
${\bf \hat B_k}$. 

The Liouville operator Eq. (\ref{eq:liouville}),(\ref{eq:heisenberg}) 
for an open quantum system can be partitioned into a
unitary part ${\cal L}_H$ and a dissipative part ${\cal L}_D$ \cite{lindblad76}: 
\begin{equation}
{\cal L}={\cal L}_H+{\cal L}_D~~~.
\label{eq:lioville2}
\end{equation}
The unitary part is generated by the Hamiltonian:
$\bf \hat H$:
\begin{equation} 
{\cal L }_H^*({\bf \hat A}) =i[{\bf \hat H},{\bf \hat A}]~~.
\label{eq:unitary1}  
\end{equation}
The condition  for a set of operators to be closed under ${\cal L}_H^*$ 
have been  well studied \cite{yoram}.
If the Hamiltonian can be decomposed to:
\begin{equation}
{\bf \hat H} = \sum _{\bf j} h_{\bf j} {\bf \hat B_j} ~~~,
\label{eq:lhamil}
\end{equation} 
and the set ${\bf \hat B_k}$ forms a Lie algebra \cite{paldus88,wybourne} i.e. 
$[{\bf \hat B_i},{\bf \hat B_j}]= \sum_{\bf k} C_{\bf ij}^{\bf k} {\bf \hat B_k}$
(the coefficients $C_{\bf ij}^{\bf k}$ are the structure factors of the Lie algebra),
then the set is closed under ${\cal L}_H^*$.

For the dissipative Liouville operator ${\cal L}_D$,  Lindblad's 
form is used \cite{lindblad76}:
\begin{eqnarray}
{\cal L}_{D}^* (\bf \hat A)  ~~=~~
\Large 
\sum_{j}~\left(
{\bf \hat F}_{\rm j} 
{\bf \hat A} {\bf \hat F}_{\rm j}^{\dagger} -  \frac{ 1}{ 2 } ( \bf \hat F_{\rm j}
{\bf \hat F}_{\rm j}^{\dagger} {\bf \hat A}~+~{\bf \hat A} {\bf \hat F}_{\rm j}
{\bf \hat F}_{\rm j}^{\dagger}) \right)~~~,
\label{eq:Lindblad2}
\end{eqnarray}
\normalsize
where $\bf \hat F_{\rm j}$ are operators from the Hilbert space of the system.
The conditions for which the set ${\bf \hat B}_i$ is closed to ${\cal L}_D^*$
have not been well established. Nevertheless in the present  studied example 
such a set has been found.

\subsection{Energy balance}

The energy balance of the working medium is followed by 
the changes in time to the expectation value of the Hamiltonian operator. 
For a working medium composed of a gas of interacting particles
the Hamiltonian is described as:
\begin{equation}
{\bf \hat H} ~~=~~  {\bf \hat H_{ext}} ~+~  {\bf \hat H_{int}}~~~.
\label{eq:hamil}
\end{equation} 
${\bf \hat H_{ext}}= \omega \sum_i {\bf \hat H_i} $ is the sum 
of single particle Hamiltonians, where $\omega= \omega(t)$ 
is the time dependent external field. It therefore constitutes the external
control of the engine's operation cycle. $ {\bf \hat H_{int}}$
represents the uncontrolled inter-particle interaction part.

The existence of the interaction term in the Hamiltonian means that
the external field only partly controls the energy of the system.
One can distinguish two cases, the first is when
the two parts of the Hamiltonian $\bf \hat H_{ext}$ and $\bf \hat H_{int}$
commute. The other case occurs when $[{\bf \hat H_{ext}},{\bf \hat H_{int}}]
\ne 0$ leads to
 $[{\bf \hat H_{int}}(t),{\bf \hat H_{int}}(t^{'})] \ne 0$,
causing important restrictions on the cycle of operation (Cf. section \ref{sec:cycle}).

Since the energy is $E= \langle {\bf \hat H} \rangle $,  the energy balance becomes 
Cf. Eq. (\ref{eq:heisenberg}):
\begin{equation}
\frac{dE}{dt}= \langle {\cal L}^*({\bf \hat H}) \rangle + 
\langle \frac{\partial {\bf \hat H}}{\partial t} \rangle ~~,
\label{eq:dedt}
\end{equation} 
Eq. (\ref{eq:dedt}) is composed of the change in time due to the explicit
time dependence of the Hamiltonian (Cf. Eq. (\ref{eq:heisenberg})
interpreted as the thermodynamic power:
\begin{equation}
{\cal P} ~~=~~ \dot \omega \sum_i \langle  {\bf \hat H_i} \rangle~~~,
\label{eq:power}
\end{equation}
where $\langle  {\bf \hat H_i} \rangle$ is the expectation value of the
single particle Hamiltonian. The accumulated work on
an engines trajectory ${\cal W} = \int{\cal P} dt$.

The heat flow represents the change in energy due to dissipation:
\begin{equation}
\dot {\cal Q}~~=~~ \langle {\cal L}_{D}^* \left( {\bf \hat H} \right)\rangle
~~=~~ \langle {\cal L}_{D}^* \left( {\bf \hat H_{\bf ext}}+{\bf \hat H_{\bf int}} \right)\rangle ~~~,
\label{eq:heatflow}
\end{equation}
(note $ {\cal L}^* ( {\bf \hat H} )~=~{\cal L}_{D}^* ( {\bf \hat H} )$
since ${\cal L}_{H}^* ( {\bf \hat H} )=0$).
Eqs. (\ref{eq:dedt}),(\ref{eq:power}) and (\ref{eq:heatflow}) 
leads to the time derivative of the first law of thermodynamics \cite{k24,gordon01,spohn79,alicki79}:
\begin{equation}
\frac{d E}{dt} ~~=~~{\cal P}+\dot {\cal Q} ~~~.
\label{eq:firstlaw}
\end{equation}

\subsection{Entropy balance}

Assuming the bath is large the entropy
production due to heat transfer from the system to the bath becomes:
\begin{equation}
{\cal D} {\cal  S}  ~~=~~\frac{\dot {\cal Q}}{T}~~~,
\label{eq:entrpord}
\end{equation}
where $T$ is the bath temperature. 

Adopting the supposition that entropy is a measure of the dispersion
of the measurement of an observable $\langle {\bf \hat A} \rangle$,
 we can label the entropy of the working medium 
according to the measurement applied i.e. ${\cal S}_{\bf \hat A}$. 
The probability of obtaining
a particular $i$th measurement outcome is: $p_i = tr\{ {\bf \hat P}_i {{{\boldsymbol{ \mathrm {  \rho}}}}} \}$
where ${\bf \hat P}_i= |i\rangle \langle i |$ 
are the projections of the $i$ th eigenvalue of the operator
$\bf \hat A$. The entropy associated with the measurement of $\bf \hat A$ becomes:
\begin{equation}
{\cal S}_{\bf \hat A} ~~=~~ - \sum_i p_i \log p_i~~~,
\label{eq:aentropy}
\end{equation}
The probabilities in Eq. (\ref{eq:aentropy}) can be obtained from the diagonal elements
of the density operator ${{{\boldsymbol{ \mathrm {  \rho}}}}}$ in the eigen-representation of ${\bf \hat A}$.
The entropy of the operator $\Op A$ that leads to minimum dispersion (\ref{eq:aentropy}), 
defines an invariant of the system termed the Von Neumann entropy \cite{wherl78}: 
\begin{equation}
{\cal S}_{VN} ~~=~~ - tr\{ {{{\boldsymbol{ \mathrm {  \rho}}}}} \log {{{\boldsymbol{ \mathrm {  \rho}}}}} \}~~~,
\label{eq:vnentropy}
\end{equation} 
${\cal S}_{\bf \hat A} \geq {\cal S}_{VN}$ for all $\bf \hat A$. 
The analysis of the energy entropy ${\cal S}_E={\cal S}_{\bf \hat H}$ of the working fluid
during the cycle of operation is a source of insight into the dynamics.
It has the property: ${\cal S}_E \geq {\cal S}_{VN}$ with equality
when the ${{{\boldsymbol{ \mathrm {  \rho}}}}}$ is diagonal in the energy representation which is true
in thermal equilibrium. Then:
\begin{equation}
{{{\boldsymbol{ \mathrm {  \rho}}}}}_{eq}~~=~~\frac{e^{-\beta{\bf \hat H}}}{Z}~~~,
\label{eq:thermal}
\end{equation}
with $\beta = 1/k_bT$ and $Z~=~tr\{e^{-\beta{\bf \hat H}}\}$, The systems temperature 
has thus become identical with the bath temperature.
When the working medium is not in thermal equilibrium,
a dynamical temperature of the working medium is defined by \cite{k94}:
\begin{equation}
T_{dyn} ~=~ \frac{\left(\frac{d E}{dt}\right)}{\left(\frac{d{\cal S}_E}{dt}\right)}~~~,
\label{eq:dytemp}
\end{equation}
and will be used to define the internal temperature of the working fluid 
(Cf. Section \ref{sec:intemp}).

\section{The quantum model}
\label{sec:model}

The following quantum model demonstrates a discrete heat engine with a cycle of operation defined by an external 
control on the Hamiltonian and by the time duration where the working medium
is in contact with the hot and cold bath.
The model studied is a particular realization of the general framework of section 
\ref{sec:formal}. First the generators of the motion ${\cal L}_H $ and ${\cal L}_D $
are derived leading to equations of motion. These equations of motion are then
solved for each of the branches thus constructing the operating cycle.

\subsection{The equations of motion }

The generators of the equations of motion are the Hamiltonian for
the unitary evolution and ${\cal L}_D$ for the dissipative part (Cf. Eq. (\ref{eq:lioville2})).

\subsubsection{The Hamiltonian}

The single particle Hamiltonian is chosen to be proportional
to the polarization of a two-level-system (TLS):
${\boldsymbol{\mathrm{\hat{\sigma}}}}_z^j$, which can be realized
as an ensemble of spins in an external time dependent magnetic field.
The operators ${\boldsymbol{\mathrm{\hat{\sigma}}}}_z,
       {\boldsymbol{\mathrm{\hat{\sigma}}}}_x,
     { \boldsymbol{\mathrm{\hat{\sigma}}}}_y$ are the Pauli matrices.
For this system, the external Hamiltonian, Eq. (\ref{eq:power}) becomes:
\begin{equation}
{\bf \hat H_{ext}} ~~=~~2^{-3/2} \omega(t) 
\left({\boldsymbol{\mathrm{\hat{\sigma}}}}_z^1 
\otimes {\bf \hat I^2}
+
{\bf \hat I^1} \otimes {{{{\boldsymbol{ \mathrm {  \sigma}}}}}_z^2}  \right)~~~,
\label{eq:hext1}
\end{equation}
and the external control field $\omega(t)$ is chosen to be in the  $z$ direction. 
The uncontrolled interaction Hamiltonian is chosen to be restricted to coupling 
of pairs of spin atoms. Therefore  the working fluid consists of noninteracting pairs
of TLS's.  For simplicity, a single pair can be considered.
The thermodynamics of $M$ pairs then follows by introducing a trivial scale factor. 
Accordingly the uncontrolled part is:
\begin{equation}
{\bf \hat H_{int}} ~~=~~2^{-3/2} J \left({ {\boldsymbol{\mathrm{\hat
{\sigma}}}}_x^1} \otimes { {\boldsymbol{\mathrm{\hat{\sigma}}}}_x^2} -
{{\boldsymbol{\mathrm{\hat{\sigma}}}}_y^1}\otimes {\boldsymbol
{\mathrm{\hat{\sigma}}}}_y^2 ~~~.
  \right)
\label{eq:interaction}
\end{equation}
$J$ scales the strength of the interaction. When $J \rightarrow 0$,
the model represents a working medium with noninteracting atoms \cite{feldmann96}.
The interaction term, Eq. (\ref{eq:interaction}), defines a correlation
energy between the two spins in the $x$ and $y$ directions. As a result,
the interaction Hamiltonian does not commute with the external Hamiltonian
Eq. (\ref{eq:hext1}), which is chosen to be polarized in the $z$ direction.

\subsubsection{The operator algebra of the working medium}
\label{subsec:dynamics}

The maximum size of the complete operator algebra of two coupled spin systems is 16.
A minimum set of operators closed to ${\cal L}^*$ is sought which is sufficient
as the basis for describing the thermodynamical quantities.
First, a Lie algebra which is closed to the unitary evolution part is to be determined.
To generate this algebra the commutation relations between the
operators composing the Hamiltonian are evaluated ( Cf. Eq. (\ref{eq:lhamil})). Defining:
\begin{eqnarray}
\begin{array}{c}
{\bf \hat B}_{1}
\end{array} ~~=~~
\begin{array}{c}
2^{-3/2} 
\left( {\boldsymbol{\mathrm{\hat{\sigma}}}}_z^1 \otimes {\bf \hat I^2}
+
{\bf \hat I^1}  \otimes {\boldsymbol{\mathrm{\hat{\sigma}}}}_z^2
\right)
\end{array} ~~=~~\frac{1}{  \sqrt{2}}
\left(
\begin{array}{cccc}
1 & 0 & 0 & 0 \\
0 & 0 & 0 & 0 \\
0 & 0 & 0 & 0 \\
0 & 0 & 0 & -1 \\
\end{array} 
\right)~~~,
\label{matB1}
\end{eqnarray}
where the tensor product eigenstates of $\boldsymbol{\mathrm{\hat{\sigma}}}_z^1$
and $\boldsymbol{\mathrm{\hat{\sigma}}}_z^2$ are used
for the matrix representation, termed the "polarization representation".

The second operator $\bf \hat B_2$ is:
\begin{eqnarray}
\begin{array}{c}
{\bf \hat B}_{2}
\end{array} ~~=~~
\begin{array}{c}
2^{-3/2} \left( {{\boldsymbol{\mathrm{\hat{\sigma}}}}_x^1} 
\otimes {{\boldsymbol{\mathrm{\hat{\sigma}}}}_x^2} -
{{\boldsymbol{\mathrm{\hat{\sigma}}}}_y^1} 
\otimes {{\boldsymbol{\mathrm{\hat{\sigma}}}}_y^2}
\right)
\end{array} ~~=~~~\frac{1}{ \sqrt{2}}   \left(
\begin{array}{cccc}
0 & 0 & 0 & 1 \\
0 & 0 & 0 & 0 \\
0 & 0 & 0 & 0 \\
1 & 0 & 0 & 0 \\
\end{array} 
\right)~~~,
\label{matB2}
\end{eqnarray}
The commutation relation:
$ [{\bf \hat B}_{1},{\bf \hat B}_{2}] = \sqrt{2} i {\bf \hat B}_{3}$
leads to the definition of ${\bf \hat B}_{3}$
\begin{eqnarray}
\begin{array}{c}
{\bf \hat B_{3}}
\end{array} ~~=~~
\begin{array}{c}
2^{-3/2}
\left( {{\boldsymbol{\mathrm{\hat{\sigma}}}}_y^1} 
\otimes {{\boldsymbol{\mathrm{\hat{\sigma}}}}_x^2} +
{{\boldsymbol{\mathrm{\hat{\sigma}}}}_x^1} 
\otimes {{\boldsymbol{\mathrm{\hat{\sigma}}}}_y^2} \right)
\end{array} ~~=~~
~\frac{1}{  \sqrt{2}} \left(
\begin{array}{cccc}
0 & 0 & 0 & -i \\
0 & 0 & 0 & 0 \\
0 & 0 & 0 & 0 \\
i & 0 & 0 & 0 \\
\end{array} 
\right)~~~.
\label{matB3}
\end{eqnarray}

The set of operators ${\bf \hat B}_{1},{\bf \hat B}_{2}, {\bf \hat B}_{3}$
form a closed sub-algebra of the total Lie algebra of the combined system.
The Hamiltonian expressed in terms of the operators 
${\bf \hat B_1}, {\bf \hat B_2}, {\bf \hat B_3}$ becomes:
\begin{eqnarray}
\begin{array}{c}
{ \bf \hat H }
\end{array} ~~=~~
 \begin{array}{c}
\omega {\bf \hat B_{1}}+\rm J {\bf \hat B_{2}}
\end{array}  ~~=~~~\frac{1 }{ \sqrt{2}}   \left(
\begin{array}{cccc}
\omega & 0 & 0 & J \\
0 & 0 & 0 & 0 \\
0 & 0 & 0 & 0 \\
J & 0 & 0 & -{\omega}\\
\end{array} 
\right)~~~.
\label{matHP}
\end{eqnarray}
All the three operators are Hermitian, and orthogonal (Cf. Eq. (\ref{eq:ortho})~).
Table (I) summarizes the commutation relations of this set of operators.
\begin{table}
\begin{center}
\caption{\rm Multiplication table of the commutation relations $[{\bf \hat X},{\bf \hat Y}]$
\rm of the operators
$\bf \hat B_l$ \rm  between themselves and with the Hamiltonian.} 
\vspace{0.3cm}
\begin{tabular}{|c|c|c|c|}
\hline
${\bf \hat X} \backslash {\bf \hat Y}$& ${\bf \hat B_1}$&$ {\bf \hat B_2}$&$\bf \hat B_3$ \\
\hline 
${\bf \hat B_1}$&$0$&$i \sqrt{2}{ \bf B_3}$&$-i \sqrt{2} {\bf \hat B_2}$ \\
\hline
${\bf \hat B_2}$&$-i \sqrt{2} {\bf \hat B_3}$&$0$&$i \sqrt{2} {\bf \hat B_1}$ \\
\hline
${\bf \hat B_3}$&$i \sqrt{2} {\bf \hat B_2}$&$-i \sqrt{2} {\bf \hat  B_1}$&$0$ \\
\hline
$\bf \hat H$&$-i \sqrt{2} \rm J{\bf \hat B_3}$&$i \sqrt{2} \omega {\bf \hat B_3}$&
$i \sqrt{2} \rm J{\bf \hat B_1}-i \sqrt{2}  
\omega {\bf \hat B_2}$ \\  
\hline
\end{tabular}
\end{center}
\vspace{0.3cm}
\label{tab:multB13}
\end{table} 

The commutation relations of the set of  $\bf \hat B_k$ operators 
define the SU(2) group and 
are isomorphic to the angular momentum commutation relations 
by the transformation ${\bf \hat B_k}~ \rightarrow ~{\bf \hat J_k}$.
${\bf \hat B_1},{\bf \hat B_2} , {\bf \hat B_3}$ can be identified as the generators of rotations
around the $z,x$ and $y$ axes respectively.
This representation allows to express the expectation values in a Cartesian three 
dimensional space ( See  Fig. \ref{fig:cyctraj}) .

\subsubsection{The generators of the dissipative dynamics}                                                                               
  
The dissipative part of the dynamics is responsible for the approach to
thermal equilibrium when the working medium is in contact with the hot/cold baths. 
The choice of Lindblad's form in Eq. (\ref{eq:Lindblad2}) guarantees the positivity
of the evolution \cite{lindblad76}. The operators $\bf \hat F_{\rm j}$ 
which lead to thermal equilibrium are constructed from
the transition operators between the energy eigenstates.
Diagonalizing the Hamiltonian (\ref{eq:hamil}) leads to the set of energy eigenvalues and eigenstates:
\begin{eqnarray}
\epsilon_1= -\frac{ \Omega }{ \sqrt{2}},~~~ \epsilon_2= 0,~~~ \epsilon_3= 0,~~~ 
\epsilon_4= \frac{ \Omega}{ \sqrt{2} }~~~,
\label{ENEHA}
\end{eqnarray}
where $\Omega = \sqrt{\omega^2+J^2}$.
The method of construction of $\bf \hat F_{\rm j}$ is based on identifying the operators 
with the raising and lowering operators in the energy frame. For example,
${\bf \hat F_{1}}= \sqrt{k \downarrow} |2 \rangle \langle 1|$ or
${\bf \hat F_{2}}= \sqrt{k \uparrow} |1 \rangle \langle 2|$.
The bath temperature enters through the detailed balance relation \cite{feldmann96,geva0}
\begin{equation}
\frac{k \uparrow}{k \downarrow}~~=~~e^{-\beta \frac{ \Omega }{ \sqrt{2} }}~~~.
\label{eq:detailed}
\end{equation}
The operators $\bf \hat F_{\rm j}$ constructed in the  energy frame  are then 
transformed into the polarization representation.  
The details are described in Appendix B.

Substituting the $\bf \hat B_i$ operators into ${ \cal L}_D $, Eq. (\ref{eq:Lindblad2}), 
one gets: 
\begin{eqnarray}
\begin{array}{l}
{ \cal L}_{D} ({\bf \hat B_1})  ~~=~~-\Gamma ({\bf \hat B_1} + 
\frac{ \omega }{ \sqrt{2} \Omega } 
\frac{{k \downarrow}-{k \uparrow} }{ \Gamma}~{\bf \hat I}) \\
{ \cal L}_{D} ({\bf \hat B_2})  ~~=~~-\Gamma ({\bf \hat B_2} + 
\frac{\rm J }{  \sqrt{2} \Omega } 
\frac{{k \downarrow}-{k \uparrow} }{ \Gamma}~{\bf \hat I}) \\
{ \cal L}_{D} ({\bf \hat B_3})  ~~=~~-\Gamma ({\bf \hat B_3} )~~~,
\end{array} 
\label{LindbB3}
\end{eqnarray}
where $\Gamma~=~{k \downarrow}+{k \uparrow}$.

From Eq. (\ref{LindbB3}) the set of $\{\bf \hat B \}$ operators and the identity operator $\bf \hat I$
are invariant to the application of the dissipative operator ${ \cal L}_{D}$
which leads to equilibration. 

The interaction of the working medium with the bath can also be elastic. 
These encounters will scramble the phase conjugate to the energy of the system
and are classified as pure dephasing ($T_2$) (Cf. Eq. (\ref{eq:ldynm}).
In Lindblad's formulation the dissipative generator of elastic encounters
is described as:
\begin{equation}
{\cal L}_{D^e}^*( {\bf \hat A} ) ~~=~~ -\gamma [ {\bf \hat H},[{\bf \hat H},{\bf \hat A}]]~~~.
\label{eq:dissip}
\end{equation}
The elastic property is equivalent to ${\cal L}_{D^e}^*( {\bf \hat H} )=0$.
Moreover the set ${\bf \hat B_i}$ which is closed to the commutation relation with $\bf \hat H$
is also closed to ${\cal L}_{D^e}^*$. 

To summarize the set ${\bf \hat B_1},{\bf \hat B_2},{\bf \hat B_3}$ and $\bf \hat I$ is closed under the
operation of ${\cal L}^*={\cal L}_H^*+{\cal L}_D^*+{\cal L}_{D^e}^*$.
Gathering together the various contributions leads to the explicit form of the equation of motion:
\begin{eqnarray}
\frac{d}{dt}
\left( \begin{array}{c}
\langle  {\bf \hat B_{1}} \rangle \\
\langle  {\bf \hat B_{2}} \rangle \\
\langle  {\bf \hat B_{3}} \rangle \\
\end{array} \right)=
\left(
\begin{array}{ccc}
-\Gamma-2 \gamma J^2 & -2 \gamma J \omega &\sqrt{2} J \\
-2 \gamma \omega J &-\Gamma-2 \gamma \omega^2&-\sqrt{2} \omega \\
-\sqrt{2} J&\sqrt{2} \omega&-\Gamma-2 \gamma \Omega^2 \\
\end{array} 
\right)
\left(\begin{array}{c}
\langle{\bf \hat B_{1}} \rangle \\
\langle{\bf \hat B_{2}}\rangle \\
\langle{\bf \hat B_{3}} \rangle \\
\end{array} \right)-
\left(
\begin{array}{c}
\frac{ \omega }{ \sqrt{2} \Omega }( {{k \downarrow}-{k \uparrow}})      \\
\frac{ J }{  \sqrt{2} \Omega }( {{k \downarrow}-{k \uparrow}})          \\
{0} \\
\end{array} \right)
\label{expmowithg}  
\end{eqnarray}
or in vector form where $b_k=\langle{\bf \hat B_{k}} \rangle$:
\begin{eqnarray}
\frac{d} {dt}  {\vec {\bf b} }  ~~=~~ 
{\cal B} 
{\vec {\bf b} } - {\vec  {\bf c}}~~~.
\label{eq:vecbprg}
\end{eqnarray}

\subsection{Integrating the equations of motion}

The thermodynamical observables require the solution of the equations of motion
on all branches of the engine. The field values $\omega$
are time independent on the {\it isochores}  thus allowing a closed form solution. 
$\omega$ changes with time on the {\it adiabats} therefore solving the equation of motion
either  requires a numerical solution
or finding  a particular solution based on an explicit time dependence of $\omega$.

\subsubsection{Solving the equations of motion on the {\it isochores}.}

On the {\em isochores} the coefficients in  Eq. (\ref{eq:vecbprg}) are time independent.
A solution is found by diagonalizing the ${\cal B}$ matrix leading to 
the eigenvalues:
$-\Gamma -i \sqrt{2} \Omega~-~2 \gamma \Omega^2,~-\Gamma~$ and 
$ -\Gamma+i \sqrt{2}  \Omega~-~2 \gamma \Omega^2 $. 
The diagonalization enables to perform in closed form the exponentiation of $e^{{\cal B^{'}}\Delta t}$  
obtaining the propagator of the  working medium operators $ { \cal U}(\Delta t) $.
\begin{eqnarray}
\nonumber
\begin{array}{c}
{ \cal U}(\Delta t)=
\end{array}
\begin{array}{c}
 { \cal R}       
\end{array}
\left(\begin{array}{ccc}
e^{-({\Gamma +i \sqrt{2}  \Omega+2 \gamma \Omega^2})\Delta t}&0&0 \\
0&e^{(-\Gamma \Delta t)}&0 \\
0&0&e^{-({\Gamma-i \sqrt{2} \Omega+2 \gamma \Omega^2})\Delta t} \\
\end{array} 
\right)
\begin{array}{c}
 { \cal R}^{-1}
\end{array}
\end{eqnarray}
where:
\begin{eqnarray}
 \begin{array}{c}
{\cal R}
\end{array} ~~=~~
\left(\begin{array}{ccc}
{{iJ}/{\sqrt{2} \Omega}}&{\omega/\Omega}  & -{{iJ}/{\sqrt{2} \Omega}} \\
-{{i \omega}/{\sqrt{2} \Omega}}&  {J/\Omega} 
&{{i \omega}/{\sqrt{2} \Omega}} \\
 {1/\sqrt{2}} &0 & {1/\sqrt{2}}  \\
\end{array} \right)~~~,
\label{gamvec1} 
\end{eqnarray}
leading to the final result:
\begin{eqnarray}
\begin{array}{c}
{ \cal U}(\Delta t)=
\end{array}
\begin{array}{c}
\exp{-(\Gamma+2 \gamma \Omega^2) \Delta t}
\end{array}
\left(
\begin{array}{ccc}
 \frac{X \omega^2+c{J^2}}{{\Omega}^2}&
\frac{{\omega}J(X-c)}{{\Omega}^2}
 &\frac{Js}{\Omega} \\
\frac{{\omega}J(X-c)}{{\Omega}^2}&\frac{X J^2+c{\omega}^2}{{\Omega}^2} &
\frac{{-\omega}s}{ \Omega}\\
 -\frac{Js}{ \Omega} &\frac{{\omega}s}{\Omega}& c \\
\end{array} 
\right)~~~,
\label{PropagTNew}
\end{eqnarray}
where  $X=\exp({2 \gamma \Omega^2 \Delta  t})$,
$c=\cos(\sqrt{2} \Omega \Delta t)$ and $ s=\sin(\sqrt{2} \Omega \Delta t)$. 
The solution of Eq. (\ref{expmowithg}) then becomes:
\begin{eqnarray}
\vec { \bf b } (t+ \Delta t)~~=~~ {\cal U}(\Delta t) ( \vec {\bf b} (t) - 
 \vec {\bf b^{eq}}) + \vec {\bf b^{eq}}~~~,
\label{solwgam}
\end{eqnarray}
where the equilibrium values of the operators are calculated from 
the steady state solutions of Eq. (\ref{eq:vecbprg}):
\begin{eqnarray}
\begin{array}{l}
{b_1^{eq}}~=~\langle {\bf \hat B_1^{eq}} \rangle~=~-\frac{\sqrt{2} \omega }{
 \Omega \rm Z}
\sinh({ \Omega \beta}/\sqrt{2}) ~=~-\frac{ \omega }{ \sqrt{2} \Omega }
  \frac{{k \downarrow}~-~ {k \uparrow}}  { \Gamma } \\
b_2^{eq}~=~\langle {\bf \hat B_2^{eq}} \rangle~=~-\frac{\sqrt{2} \rm J }{
 \Omega \rm Z}
\sinh({\Omega \beta}/\sqrt{2} ) ~=~-\frac{\rm J }{ \sqrt{2} \Omega }
  \frac{{k \downarrow}~-~ {k \uparrow} }{ \Gamma } \\
b_3^{eq}~=~\langle{\bf \hat  B_3^{eq}}\rangle~~=~~0~~~. 
\end{array}
\label{B3EQQ}
\end{eqnarray}
On the {\em isochores} the solution of Eq. (\ref{PropagTNew}) 
can be extended to the full duration $\tau_{h/c}$ of propagation on the hot/cold branches. 
Therefore,  $\Delta t = \tau_{h/c}$.

There are cycles of operation where the external field $\omega$ also varies
when the working medium is in contact with the hot or cold baths, 
for example the Carnot cycle \cite{geva1}.
For such cycles the equation of motion can be solved
by decomposing these branches into small segments of duration $\Delta t$.
Then  Eq. (\ref{solwgam}) can be used as an approximate to the short time propagator.

\subsection{Propagation of the observables on the {\em adiabats}} 
\label{subsec:adiabr}
 
The equations of motion on the {\em adiabats} have explicit time dependence.
To overcome this difficulty two approaches are followed. The first is based on
decomposing the evolution to short time segments
and using a short time approximation to solve the equations of motion. 
The second approach is based on finding a particular time dependence form of $\omega(t)$ which
allows an analytic solution.

\subsubsection{Short time approximation}
\label{subsubsec:shorttime} 
 
For the {\em adiabatic}  branches the working medium 
is decoupled from the baths so that the time propagation 
is unitary. Eq. (\ref{expmowithg}) thus simplifies to:
\begin{eqnarray}
\frac{d}{dt}\left( \begin{array}{c}
{ b_{1}} \\
{ b_{2}} \\
{ b_{3}} \\
\end{array} \right) ~~=~~ \left(
\begin{array}{ccc}
0&0&\sqrt{2} J \\
0&0&-\sqrt{2} \omega(t) \\
-\sqrt{2} J&\sqrt{2} \omega(t)&0 \\
\end{array} 
\right)
\left(\begin{array}{c}
~b_{1} \\
~b_{2} \\
~b_{3} \\
\end{array} \right)~~~.
\label{eqmot1VN} 
\end{eqnarray}
Or in the vector form: $\frac{d}{dt} {\vec {\bf  b}} ~~=~~\tilde L(t) {\vec  {\bf b}}$.
Since the matrix $\tilde L(t)$ is time dependent
the propagation is broken into short time segments $\Delta t$,
reflecting the fact that  $[\tilde L (t),\tilde L(t')] \neq 0$,   
\begin{equation}
{\vec {\bf b}} (t) ~~=~~ \prod_{j=1}^{N} \exp 
\left( \int_{(j-1)\Delta t}^{j\Delta t} {\tilde L}(t')dt' \right) {\vec 
{\bf b}}(0)~~~,
\label{eq:propb}
\end{equation}  
where $N \Delta t = t$.     
Eq. (\ref{eqmot1VN})  is solved by diagonalizing the matrix $\tilde L$ 
for each time step assuming that during  the period $ \Delta t$
$\omega(t)$ is constant.
Under such conditions ${\cal U}_a(t,\Delta t) $ becomes:
( the index $a$ stands for {\em adiabat})
\begin{eqnarray}
\begin{array}{c}
{\cal U}_a(t,\Delta t)~~=~~ e^{{\tilde L}(t) \Delta t}    
\end{array}
~~=~~\left(
\begin{array}{ccc}
\frac{{\omega}^2+c{J^2}} { {\Omega}^2}&\frac{{\omega}J(1-c)}{ {\Omega}^2} 
&\frac{Js }{ \Omega} \\
\frac{{\omega}J(1-c)}  {{\Omega}^2}&\frac{J^2+c{\omega}^2} 
{{\Omega}^2} &
-\frac{\omega s} {  \Omega}\\
 -\frac{Js }{ \Omega }&\frac{{\omega}s} { \Omega}&c \\
\end{array} \right)~~~,
\label{propag}
\end{eqnarray}
which becomes the short time propagator for the {\em adiabats} 
from time $t$ to $t+\Delta t$.

\subsubsection{An analytical solution on the {\em adiabats}}
\label{subsubsec:analytsol} 

The analytic solution for the propagator on the {\em adiabats}  is based on the Lie group
structure of the $\{\bf \hat B \}$ operators. The solution is based on the unitary evolution
operator $\bf \hat U (t)$ which for explicitly time dependent Hamiltonians is 
obtained from the Schr\"odinger equation:
\begin{eqnarray}
-i\frac{d}{dt}
{{\bf \hat  U} (t)}  ~=~ 
{\bf \hat H} (t){\bf \hat U} (t),   ~~~~~~~~  {\bf \hat U} (0)={\bf \hat I }~~~.
\label{weinorde} 
\end{eqnarray}
The propagated set of operators becomes:
\begin{eqnarray}
{\vec {\bf \hat B}} (t)    ~=~  {\bf \hat U} (t) \vec {\bf \hat B} (0) {\bf \hat U}^{\dagger} (t)~=~
  {\cal U}_a(t)    {\vec {\bf \hat B}}  (0) ~~~,     
\label{sandwich}
\end{eqnarray}
and is related to the super-evolution operator ${\cal U}_a(t)$.
Based on the group structure Wei and Norman, \cite{weinorman63} 
constructed a solution to Eq. (\ref{weinorde}) for any
operator ${\bf \hat H}$ which can be written as a linear combination
of the operators in the closed Lie algebra
${\bf \hat H}  (t)  ~=~ \sum_{j=1}^m h_j(t){\bf \hat B}_i,  $,
where the $h_i(t)$ are scalar functions of $t$, ( Cf. Eq. (\ref{eq:lhamil})). 
In such a case the unitary evolution operator  ${\bf \hat U} (t)$ 
can be represented in the product form:
\begin{eqnarray}
{\bf \hat U} (t)  ~=~ 
\prod_{k=1}^{m} \exp(\alpha_k(t){\bf \hat B} _k) ~~~. 
\label{weinorde2} 
\end{eqnarray} 
The product form replaces the time dependent operator equation  (\ref{propag}) 
with a set of scalar differential
equations for the functions $\alpha_k(t)$.
As has been shown in \ref{subsec:dynamics}, three $ \bf \hat B_k$ operators form a closed
Lie Algebra. Writing the unitary evolution operator explicitly leads to:
\begin{eqnarray} 
{\bf \hat U} (t)   ~=~  
 \exp(i\frac{ \alpha_1(t) }{ \sqrt{2} } {\bf \hat B_1} ) \exp(i \frac{\alpha_2(t)
}{ \sqrt{2}}{\bf \hat B_2} )
\exp(i \frac{\alpha_3(t) }{\sqrt{2}}{\bf \hat B_3} )    
\label{weinorde3}  
\end{eqnarray}  
The $\sqrt{2}$ factor is introduced for technical reasons.
Based on the group structure \cite{weinorman63} Eq. (\ref{weinorde}) leads to the following 
set of differential equations has to be solved:
\begin{eqnarray}
\dot \alpha_1= \sqrt{2} \omega(t)+ \sqrt{2} J (\frac{\sin(\alpha_1)
\sin(\alpha_2) }{ \cos( \alpha_2)})~;~~ 
\dot \alpha_2=  \sqrt{2} J \cos(\alpha_1)~;~~
\dot \alpha_3= \frac{ \sqrt{2} J \sin(\alpha_1)}{ \cos(\alpha_2) }~~~. 
\label{mateq3}   
\end{eqnarray}
Using Eq.  (\ref{sandwich}) the propagator  $ {\cal U}_a(t)$
is evaluated explicitly in terms of the coefficients $\alpha$: 
\begin{eqnarray}
 {\cal U}_a(t)      ~~=~~ \left(
\begin{array}{ccc}
c_2c_3 & -s_3c_1+c_3s_2s_1&c_3s_2c_1+s_3s_1 \\
c_2s_3   &c_3c_1+s_3s_2s_1& s_3s_2c_1-c_3s_1 \\
 -s_2 &   c_2s_1 &c_2c_1 \\
\end{array} 
\right)~~~,
\label{propan}
\end{eqnarray} 
where:
$s_1=\sin(\alpha_1),~ s_2=\sin(\alpha_2),~ s_3=\sin(\alpha_3)$,~
$c_1=\cos(\alpha_1),~ c_2=\cos(\alpha_2),~ c_3=\cos(\alpha_3) $. 
 
The problem of obtaining a closed form solution for the propagator $ {\cal U}_a(t)$
has been transformed in to finding the solution of three coupled differential equations
Eq. (\ref{mateq3} ) which depend on $\omega(t)$. A general solution has not been found
but by choosing a particular functional form for $\omega(t)$ a closed form solution has been 
obtained.

\subsubsection{The Explicit Solution for $\alpha$}
\label{subsec:explicit}

To facilitate the solution of  Eq. (\ref{mateq3}), a particular form of $\omega (t)$ is chosen:
\begin{eqnarray}
\omega(t)= \frac{ \dot \alpha_1 }{ \sqrt{2} }-J \frac{\sin(\alpha_1)\sin(\alpha_2)}{
 \cos(\alpha_2) }~~~.
\label{omegag}
\end{eqnarray}
Two auxiliary functions are defined, $u(t)$ and $v(t)$:
\begin{eqnarray}
u(t)=-J^2t^2+\sqrt{2}rJt;~~v(t)=r-\sqrt{2}Jt~~~~.      
\label{auxiliary}
\end{eqnarray}
$r$ is a constant which restricts the product $Jt$: $\{~~0 < r < 1;~Jt< \sqrt{2} r \}$.
In terms of $u(t)$ and $v(t)$, the solutions of Eq. (\ref{mateq3}) become: 
\begin{eqnarray}
\alpha_1 ~~=~~ \arccos{\left(\frac{ 1 }{ \sqrt{1+2u} } \right)}~~~.
\label{solmat1} 
\end{eqnarray}
\begin{eqnarray}
\alpha_2~~=~~ \arcsin{ \left(\frac{1 }{ 1+r^2}(r \sqrt{1+2u}-v) \right)}~~~.
\label{solmat12} 
\end{eqnarray} 
\begin{eqnarray}   
\nonumber
\alpha_3~~=~~-\frac{r }{ 2} \ln(2 \sqrt{4u^2+2u}+4u+1) 
\end{eqnarray} 
\begin{eqnarray} 
\nonumber
-\frac{\sqrt{1-r^2}}{ 2} \{ \arcsin \left( \frac{2r^2(1-r^2) }{ 2u+1-r^2 } 
+1-2r^2 \right)-\frac{ \pi }{ 2} \}
-\{ \arcsin{ \left(\frac{ v }{ r } \right)}-\frac{\pi }{ 2} \}
\end{eqnarray}  
\begin{eqnarray} 
-\frac{\sqrt{1-r^2}}{ 2 }
\{ \arcsin{ \left( \frac{1 }{ r }[1- \frac{1-r^2 }{ 1+v}] \right)} +
\arcsin{ \left( \frac{1 }{ r }[1- \frac{1-r^2 }{ 1-v}] \right)} \}~~~. 
\label{solmat13}
\end{eqnarray} 
For $t=0$, ${\bf \hat U} ={\bf \hat I}$,
therefore $\alpha_{1}(0)= 0,~ \alpha_{2}(0)= 0,~ \alpha_{3}(0)=0$ 
which is consistent with Eq. (\ref{solmat1}),(\ref{solmat12}) and (\ref{solmat13}).

\begin{figure}[tb]
\vspace{-0.66cm}
\hspace{3.cm}
\psfig{figure=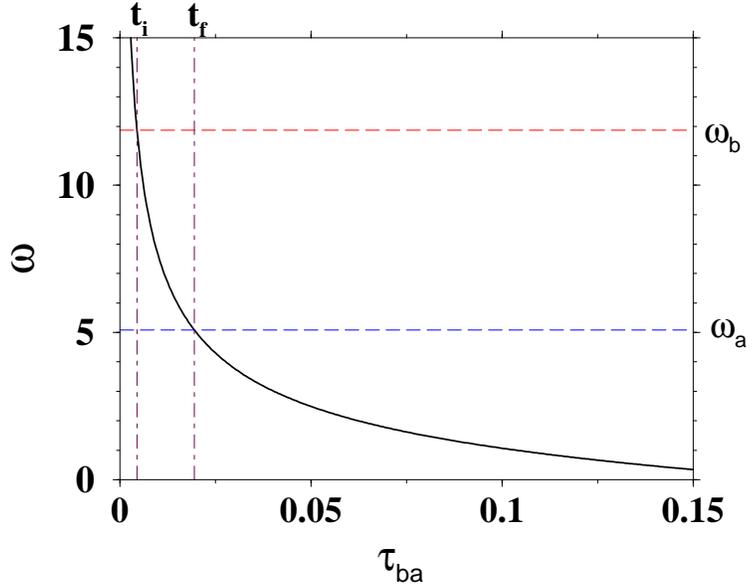,width=0.6\textwidth}
\vspace{0.5cm}
\caption{The external field $\omega$ as a function of time on the {\em adiabats} 
corresponding to the function Eq. (\ref{omegaex}) for which an analytic solution exist.
Indicated are the values of the initial and the final time and of the corresponding $\omega$
which are used to construct the cycle of operation.
Notice the singularity at $t=0$.
}
\label{fig:omega} 
\end{figure}

Introducing into  Eqs. (\ref{omegag}) the explicit functional forms of $\alpha_k$,
$\omega(t)$ becomes:
\begin{eqnarray}
\omega(t)= \frac{ Jv }{\sqrt{2} (1+2u)\sqrt{u} }-
J \frac{ \sqrt{2} \sqrt{u} (r \sqrt{1+2u}-v)
 }{ \sqrt{1+2u}( \sqrt{1+2u}+rv)    }~~~.
\label{omegaex}
\end{eqnarray}
At $t=0$, $\omega$  is singular. Since the engine operates between two
finite values of $\omega$ a corresponding time segment is chosen
which does not include the singularity at $t=0$ (Cf. Fig. \ref{fig:omega}). 
Using the group property of ${ \cal U}_a(t)$,  i.e.
${\cal U}_a(t_1){ \cal U}_a(t_2)={ \cal U}_a(t_1+t_2)$ the propagation is carried out
by changing  the origin of time, 
${ \cal U}_a(t)~=~{\cal U}^{-1}_a(t_0){ \cal U}_a(t+t_0) $ where
$t_0$ is either $t_i$ for the compression {\em adiabat} or $t_f$ for the expansion {\em adiabat}.
One should note, that ${\cal U}^{-1}_a(t)= { \cal U}^{\dagger}_a(t)$ 
but due to the explicit time dependence
${\cal U}^{-1}_a(t) \neq { \cal U}^{\dagger}_a(-t)$.

\section{Reconstruction  of ${{{\boldsymbol{ \mathrm {  \rho}}}}}_R $}
\label{recrhoR} 

The reconstruction   ${{{\boldsymbol{ \mathrm {  \rho}}}}}$ is designed to describe the state of the working medium
from its initial state to equilibrium. As was analyzed in the previous section,
the set of operators ${\bf \hat B}_1,{\bf \hat B}_2,{\bf \hat B}_3, {\bf \hat I}$ are sufficient to describe
the energy changes during the cycle of operation of the engine.
Is this set sufficient to reconstruct the density operator?

In equilibrium ${{{\boldsymbol{ \mathrm {  \rho}}}}}^{eq}$ is diagonal in the energy representation, From 
the  eigenvalues of the Hamiltonian Eq. (\ref{ENEHA}), ${{{\boldsymbol{ \mathrm {  \rho}}}}}^{eq}$
in the energy picture becomes:
\begin{eqnarray}
\begin{array}{c}
\hat \rho^{eq}~~=~~
\end{array}
\left(\begin{array}{cccc}
\frac{e^{ \Omega \beta/ \sqrt{2} }} { Z} &0&0&0 \\
0&\frac{1 }{ Z}&0 &0\\
0&0&\frac{1 }{ Z}  &0\\
0&0 & 0&\frac{e^{- \Omega \beta / \sqrt{2}}} { Z}    \\
\end{array} 
\right)~~~,
\label{DIARHO}
\end{eqnarray}
where
\begin{eqnarray}
Z~~=~~\exp{-( \Omega \beta / \sqrt{2})}~+~2~+~\exp{( \Omega \beta / \sqrt{2})}~=~
 \frac{ {k \uparrow} }{ {k \downarrow}}~+2+  
\frac{ {k \downarrow} }{ {k \uparrow}}~=~ \frac{\Gamma^2 }{
 {k \downarrow} {k \uparrow}}~~~.
\label{part1Z}
\end{eqnarray}
By inspection, the diagonal elements of the equilibrium density operator are seen to be 
defined by three independent variables. The energy expectation accounts for one
variable.
The expectation value of $\Op B_3$ has no diagonal elements in the
energy representation therefore two  additional operators are required to 
facilitate a reproduction of  $\Op \rho_R$.
\begin{eqnarray}
\begin{array}{c}
{\bf \hat B}_{4}
\end{array} ~~=~~
\begin{array}{c}
2^{-3/2} \left(  {\boldsymbol{\mathrm{\hat{\sigma}}}}_z^1 
\otimes {\bf \hat I^2} 
-
{\bf \hat I^1}  \otimes {\boldsymbol{\mathrm{\hat{\sigma}}}}_z^2 \right)
\end{array} ~~=~
~~\frac{1 }{ \sqrt{2}}  \left(
\begin{array}{cccc}
0 & 0 & 0 & 0 \\
0 & 1 & 0 & 0 \\
0 & 0 & -1 & 0 \\
0 & 0 & 0 & 0 \\
\end{array} 
\right)~~~,
\label{matB4}
\end{eqnarray}
and
\begin{eqnarray}
{\bf \hat B}_{5}
~=~
\frac{1 }{ 2}
{\boldsymbol{\mathrm{\hat{\sigma}}}}_z^1 \otimes
{\boldsymbol{\mathrm{\hat{\sigma}}}}_z^2
~=~\frac{1 }{ 2 }  
\left(
\begin{array}{cccc}
1 & 0 & 0 & 0 \\
0 & -1 & 0 & 0 \\
0 & 0 & -1 & 0 \\
0 & 0 & 0 & 1 \\
\end{array} 
\right)~~~.
\label{matB5}
\end{eqnarray}
Since both $\Op B_4$ and $\Op B_5$ commute with the 
Hamiltonian, they undergo only dissipative dynamics but they are uninfluenced by the dephasing
generated by $ ~{\cal L}_{D}^*$:
\begin{equation}
\dot {\bf \hat B_{4}}~~=~~
-\Gamma{\bf \hat B_{4}}
\label{eq:B4mot}
\end{equation} 
with the solution:
\begin{equation}
{\bf \hat B_{4}}(t)~~=~~{\bf \hat B_{4}}(0)\exp{(-\Gamma t) }~~~,
\label{eq:B4sol}
\end{equation} 
and $\langle {\bf \hat B_{4}^{eq}}\rangle =0$.
The equation of motion of $\bf \hat B_5$ is:
\begin{equation}
\nonumber
\dot {\bf \hat B_{5}}~~=~~
-2 \Gamma {\bf \hat B_{5}}- \sqrt{2}\frac{\omega }{ \Omega}
( {k \downarrow}~-~ {k \uparrow}) {\bf \hat B_{1}}-
 \sqrt{2}\frac{\rm J }{ \Omega}
( {k \downarrow}~-~ {k \uparrow}) {\bf \hat B_{2}}~~~=~~~   
\end{equation} 
\begin{equation}
-2 \Gamma{\bf \hat B_{5}}+~~2\Gamma \langle {\bf \hat B_{1}^{eq}} \rangle
\bf \hat B_{1}~~+~~~2\Gamma \langle {\bf \hat B_{2}^{eq}}\rangle
\bf \hat B_{2} ~~~. 
\label{eq:B5mot}
\end{equation} 
At equilibrium, $\dot {\bf \hat B_{5}} =0$, and then
$\langle {\bf \hat B_{5}^{eq}} \rangle ~~=~~
({\bf \hat B_{1}^{eq}})^2~~+~~({\bf \hat B_{2}^{eq}})^2  $,
a result which can be verified by computing 
$\langle {\bf \hat B_{5}^{eq}} \rangle ~~=~~
~{\rm tr} \{{\Op \rho^{eq}} {\bf \hat B_{5} } \}$.
Eq. (\ref{eq:B5mot}) is a linear first order inhomogeneous equation for
${\bf \hat B_{5}}$ depending  on the time dependence 
of the closed set ${\bf \hat B_{1}},{\bf \hat B_{2}},{\bf \hat B_{3}}$, Eq. (\ref{solwgam}).
Changing  Eq. (\ref{eq:B5mot}) to observables, Eq. (\ref{eq:sclar}), and by 
integrating subject to the solutions of $b_1$ and $b_2$
leads to:
\begin{equation}
\nonumber
{b_{5}}(t) ~=~\frac{2 }{ \Omega^2} \left(\omega 
(b_1(0)- b_{1}^{eq})+
J(b_{2}(0)-b_{2}^{eq})\right)
\left( \omega b_{1}^{eq}+  J~ b_{2}^{eq}\right)
(e^{-\Gamma t}-e^{-2 \Gamma t})~+~
\end{equation}
\begin{equation}
k_0
\left(
k_1~c (e^{-(\Gamma+2 \gamma \Omega^2) t})~+~k_2~s
e^{-(\Gamma+2 \gamma \Omega^2) t}
~-~
k_1 e^{-2 \Gamma t} \right)
~+~(b_{5}(0)-b_{5}^{eq})
e^{-2 \Gamma t}+ b_{5}^{eq}~~~,
\label{eq:B5solu}
\end{equation} 
where:
\begin{equation}
k_0~~=~~\frac{ 2 \Gamma (J b_{1}^{eq} -\omega   b_{2}^{eq} )  }{
 \Omega^{2}((\Gamma+2 \gamma \Omega^{2})^2+2 \Omega^2) }
\label{K05so}
\end{equation}
\begin{equation}
\nonumber
k_1~=~ \left( J(b_{1}(0)-b_{1}^{eq} )
- \omega(b_{2}(0)-b_{2}^{eq} )\right) 
( \Gamma+2 \gamma \Omega^{2})-
 \Omega (b_{3}(0)-b_{3}^{eq} )(\sqrt{2} \Omega)
\end{equation}
and
\begin{equation}
\nonumber
k_2~=~ \left( J(b_{1}(0)-b_{1}^{eq} )
- \omega(b_{2}(0)-b_{2}^{eq} )\right) 
(\sqrt{2} \Omega)~+~
 \Omega (b_{3}(0)-b_{3}^{eq} )
( \Gamma+2 \gamma \Omega^{2})~~~.
\end{equation}
Using the set of  of the five orthogonal and normalized operators 
together with the identity operator the density operator $\Op \rho_R$ is reconstructed.
Representing $\Op \rho_R$ in different basses facilitates the calculation of the different entropies.
$\Op \rho_R$ in the polarization basis becomes:
\begin{eqnarray}
\begin{array}{c}
{\Op \rho_p }
\end{array} ~~=
~~~  \left(
\begin{array}{cccc}
\frac{1 }{ 4}+ \frac{{b_1} }{ \sqrt{2}}+\frac{b_5 }{ 2} & 0 & 0 &
  \frac{{b_2} }{ \sqrt{2}} -i\frac{ {b_3} }{ \sqrt{2}}\\
0 & \frac{1 }{ 4}+ \frac{b_4 }{ \sqrt{2}}-\frac{b_5 }{ 2} & 0 & 0 \\
0 & 0 & \frac{1 }{ 4}- \frac{b_4 }{ \sqrt{2}}-\frac{b_5 }{ 2}  & 0 \\
 \frac{{b_2} }{ \sqrt{2}}+i\frac{{b_3} }{ \sqrt{2}}    & 0 & 0 & 
\frac{1 }{ 4}- \frac{b_1 }{ \sqrt{2}}+\frac{b_5 }{ 2}      \\
\end{array} 
\right)~~~.
\label{rorop1} 
\end{eqnarray}
The off diagonal elements of ${\Op \rho_p }$ are the expectation values of 
the operators:\\ 
${\Op B}_{\pm}=\frac{1}{\sqrt{2}}(\Op B_2 \pm i \Op B_3)$ which represent
the correlation between the individual spins.

The density operator $\Op \rho_{R}$ in the energy basis becomes:
\begin{eqnarray}
\begin{array}{c}
\Op \rho_{e}  
\end{array} ~~=
~~~  \left(
\begin{array}{cccc}
\frac{1 }{ 4} - \frac{ E }{ \Omega \sqrt{2} }+\frac{b_5 }{ 2} & 0 & 0 &
+ \frac{ib_3 }{ \sqrt{2}} 
-\frac{ J b_1 }{ \Omega \sqrt{2} }+
\frac{ \omega b_2 }{ \Omega \sqrt{2} }        \\
0 & \frac{1 }{ 4}+ \frac{b_4 }{ \sqrt{2}}-\frac{b_5 }{ 2} & 0 & 0 \\
0 & 0 & \frac{1 }{ 4}- \frac{b_4 }{ \sqrt{2}}-\frac{b_5 }{ 2}  & 0 \\
 -\frac{i b_3}{ \sqrt{2}} 
-\frac{J b_1 }{ \Omega \sqrt{2} }+
\frac{ \omega b_2 }{ \Omega \sqrt{2}}  & 0 & 0 & 
\frac{1 }{ 4}+ \frac{ E }{ \Omega \sqrt{2}} +\frac{b_5 }{ 2}      \\
\end{array} 
\right)~~~,
\label{rorop}
\end{eqnarray}
where $E= \omega b_1+Jb_2$.
In equilibrium, the off-diagonal elements vanish, and the 
matrix will be identical to  Eq. (\ref{DIARHO}). 
In non-equilibrium, the off-diagonal elements of $ {{{\boldsymbol{ \mathrm {  \rho}}}}}_{e}$ 
determine the "phase" Cf. Sec. \ref{sec:phase}.

To compute the Von-Neumann entropy   $ {{{\boldsymbol{ \mathrm {  \rho}}}}}_R$ is diagonalized leading to:
\begin{eqnarray}
\begin{array}{c}
\hat \rho_{vn} 
\end{array} ~~=
~~~  \left(
\begin{array}{cccc}
\frac{1 }{ 4}- \frac{D }{ \sqrt{2}}+\frac{b_5 }{ 2} & 0 & 0 &0 \\
0 & \frac{1 }{ 4}+ \frac{b_4 }{ \sqrt{2}}-\frac{b_5 }{ 2} & 0 & 0 \\
0 & 0 & \frac{1 }{ 4}- \frac{b_4}{ \sqrt{2}}-\frac{b_5}{ 2}  & 0 \\
 0 & 0 & 0 & 
\frac{1 }{ 4}+\frac{D }{ \sqrt{2}}+\frac{b_5}{ 2}      \\
\end{array} 
\right)~~~,
\label{diagvn}
\end{eqnarray}
where $D=\sqrt{b_1^2+b_2^2+b_3^2}$.

\section{Dynamical Temperature ($T_{dyn}$) on the branches }
\label{sec:intemp}

Based on  the definition of  the dynamical temperature $T_{dyn}$ in 
Eq. (\ref{eq:dytemp}),  and from Eq. (\ref{matHP}):
\begin{eqnarray}
T_{dyn}~=~ \frac{\dot \omega b_1+\omega \dot b_1
+J \dot b_2 }{-\sum \dot p_i^E(1+\log(p_i^E))}
~=~
\frac{\dot \omega b_1-\Gamma E-\frac{\Omega}{\sqrt{2}}( k \downarrow-k \uparrow)}
{-\sum \dot p_i^E(1+\log(p_i^E))}~~~,
\label{inttem3}
\end{eqnarray} 
The four probabilities $p_i^E$ are the diagonal elements of
the density operator in the energy representation  ${{{\boldsymbol{ \mathrm {  \rho}}}}}_{e}$, Eq. (\ref{rorop}). 
The derivatives of the probabilities are obtained from  
Eqs. (\ref{expmowithg}) and  (\ref{eq:B5mot}):
\begin{eqnarray}
\nonumber
\dot p_1^E=   
\frac{ -\dot \omega b_1+\Gamma E }{ \Omega  \sqrt{2}} + 
\frac{ ( k \downarrow- k \uparrow) }{ 2 } + \frac{\dot b_5 }{ 2},~~ 
\dot p_2^E=- \frac{\dot b_5 }{ 2},~~
\end{eqnarray}
\begin{eqnarray}
\dot p_3^E=- \frac{\dot b_5 }{ 2},~~
\dot p_4^E=
\frac{\dot \omega b_1- \Gamma E }{ \Omega  \sqrt{2}}- 
\frac{ ( k \downarrow-k \uparrow) }{ 2 } +  \frac{\dot b_5 }{ 2} ~~~, 
\label{derprob}
\end{eqnarray} 
where  $ {\dot b_5 }$  is obtained form Eq. (\ref{eq:B5mot}):
$\dot b_5~=~  2 \Gamma(b_1^{eq}b_1+ b_2^{eq}b_2-b_5)$.

\subsection{Dynamical temperature on the $ \it isochores$}

Evaluating the derivatives of the
probabilities in   Eqs. (\ref{derprob}) and using the fact that on the {\em isochores}
$\dot \omega$ =0, the dynamical temperature, Eq. (\ref{inttem3}), becomes:
\begin{eqnarray}
T_{dyn}~=~ \frac{\left( \Gamma E + \frac{\Omega}{ \sqrt{2}}(
k \downarrow- k \uparrow)
 \right)}{\left(
\frac{ \Gamma E }{ \Omega  \sqrt{2}}\log(p_1/p_4)+
\frac{ ( k \downarrow-k \uparrow)   }{ 2 }\log(p_1/p_4) + 
\frac{1}{2}\dot b_5 \log(p_1p_4/p_2p_3) 
\right)}~~~.
\label{inttem4}
\end{eqnarray}
A consistency check is obtained by
comparing $T_{dyn}$ for J=0 with the internal temperature of a two-level-system.
For $J=0$:
\begin{eqnarray}
T_{dyn}~= \frac{ \omega }{ \sqrt{2}
\log(\frac{1/2+  b_1/\sqrt{2}  }{1/2 -  b_1/\sqrt{2}})}~~~,    
\label{inttem5}
\end{eqnarray}
which leads to:
\begin{eqnarray}
b_1~=~-1/\sqrt{2}  \frac{ k \downarrow-k \uparrow }{
  k \downarrow+k \uparrow }~=~ -1/\sqrt{2} \tanh(
 \frac{   \omega  }{ \sqrt{2} T_{dyn}})~~~. 
\label{inttem00}
\end{eqnarray}
which is the internal temperature for a noninteracting spin system with
energy spacing $\omega/\sqrt{2}$ \cite{geva0}.

\subsection{Dynamical temperature on the $ \it adiabats$} 

On the {\em adiabats} $\dot b_4=$ and $\dot b_5=0$.            
From Eq. (\ref{inttem3}) and (\ref{derprob})
the derivatives of the probabilities on the adiabats become:
\begin{eqnarray}
\dot p_1^E~=~ -~\frac{ \dot \omega }{ \Omega \sqrt{2}};~~
\dot p_2^E~=~0;~~
\dot p_3^E~=~0;~~
\dot p_4^E~=~\frac{ \dot \omega }{ \Omega \sqrt{2}} ~~~,    
\label{derprobad}
\end{eqnarray} 
which leads to the dynamical temperature on the ${\it adiabats}$:
\begin{eqnarray}
T_{dyn}^{ad}~= \frac{ \Omega \sqrt{2}}{
\log(\frac{p_1^E }{ p_4^E}) } ~~~. 
\label{inttemad}
\end{eqnarray}

\section{The Thermodynamic quantities for the coupled spin fluid}
\label{sec:therquant}

\begin{itemize}
\item{{\bf The~ heat~ absorbed~ or~ delivered by the heat engine} \rm}

Using Eq.(\ref{eq:heatflow}) the heat ${\cal Q} \rm_{h/c}$
absorbed or delivered becomes:
\begin{eqnarray}
{\cal Q} \rm_i~=~ \rm (\exp({-\Gamma \tau_i}) - 1)(\omega_i b_1+J b_2)
\label{isogen}
\end{eqnarray}
where $i=h/c$ 
\item{ { \bf The work absorbed or delivered by the heat engine} \rm}

The power, Eq. (\ref{eq:power}), is $\langle \frac{\partial H }{ \partial t} \rangle = B_1(t) \dot \omega$.
Therefore, the work becomes:
\begin{eqnarray}
\cal W~=~ \rm \int_{\tau i}^{\tau f} b_1 \dot \omega dt~~~.
\label{adigen}
\end{eqnarray}

\item{ {\bf Entropy production. }  \rm}

The entropy production per cycle, ${\cal D S}_{cycle}$, 
created on the boundaries becomes (Cf. Eq. (\ref{eq:entrpord})):
\begin{eqnarray}
{\cal D S}_{cycle}=
-\left({\cal Q} \rm_{AB}/T_h+{\cal Q} \rm_{CD}/T_c\right)~~~.
\label{entpro}
\end{eqnarray}

\item{ $\bf Efficiency. $  \rm}

The efficiency per cycle, $\eta_{cycle}$ 
is:

\begin{eqnarray}
\eta_{cycle}=
\cal W~/{\cal Q} \rm_{AB}= \frac{ \int_{\tau i}^{\tau f} 
b_1 \dot \omega dt }{ \rm 
(\exp({-\Gamma \tau_i}) - 1)(\omega_i b_1+Jb_2 )}~~~.
\label{efficien}
\end{eqnarray}
The maximal efficiency of the engine is:
\begin{equation}
\nonumber
1-\frac{\Omega_a }{ \Omega_b}~=~1- \frac{ \sqrt{ \omega^{2}_{a}+J^{2}} }{
 \sqrt{ \omega^{2}_{b}+J^{2}}}~~~.
\end{equation}
The upper bound should be the Carnot's
efficiency,  a bound correct for all J and the fact that 
$\frac{\omega_a}{\omega_b}~>~ \frac{T_c}{T_h}$:  
\begin{eqnarray}
1- \frac{ \omega^{2}_{a}+J^{2}}{ \omega^{2}_{b}+J^{2}}~<~
1- \frac{ \omega^{2}_{a} }{ \omega^{2}_{b}}~<~1-\frac{ T^{2}_{c} }{ T^{2}_{h}}~~~.
\label{upperefic}
\end{eqnarray}

\end{itemize}

\section{The Cycle of Operation: The Otto cycle}
\label{sec:cycle}

The operation of the heat engine is determined by the properties of the working medium
and by the hot and cold baths. These properties are summarized by the 
generator of the dynamics ${\cal L}$.
The cycle of operation is defined by the external controls which
include the variation in time of the field with the periodic property 
$\omega(t)=\omega(t+\tau)$
where $\tau$ is the total cycle time synchronized with  the contact times
of the working medium with the hot and cold baths $\tau_h$ and $\tau_c$. 
In this study a specific
operating cycle composed of two branches  termed {\em isochores} where
the field is kept constant and the working medium is in contact 
with the hot/cold baths. 
In addition two branches termed {\em adiabats} where the field $\omega(t)$ varies
and the working medium is disconnected from the baths. 
This cycle is a quantum analogue of the Otto cycle. 

\begin{figure}[tb]
\vspace{0.1cm}
\hspace{3.cm}
\psfig{figure=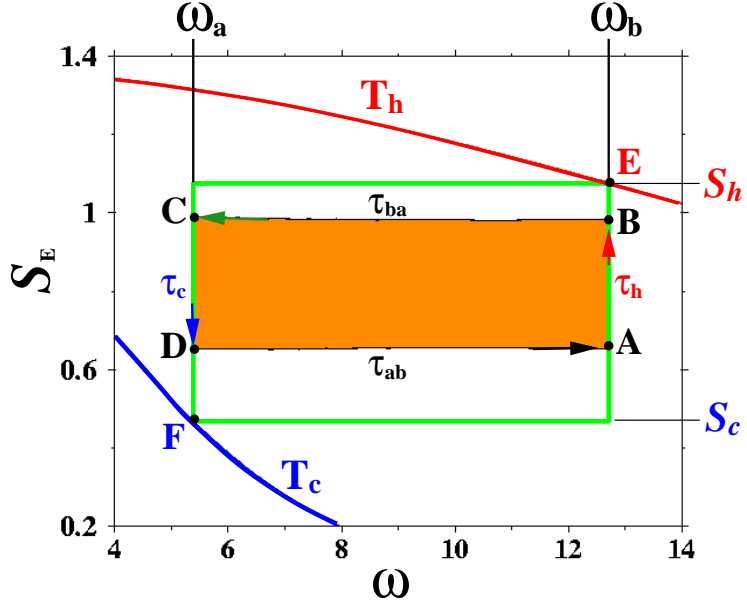,width=0.6\textwidth}
\vspace{0.5cm}
\caption{ 
The heat engine's optimal cycles in the ($\omega, {\cal S_E}$) plane. 
The upper red line indicates the energy entropy of the working medium
in equilibrium with the hot bath at temperature T$_h$ for different values of the field.
The blue line below indicates the energy entropy in equilibrium with the cold 
bath at temperature T$_c$. The cycle in green has an infinite time allocation
on all branches. It reaches the equilibrium point with the hot bath (point E)
and equilibrium point with the cold bath (point F). The inner cycle ABCD is
the optimal cycle with the optimal time allocation on all branches. calculated numerically
for a linear $\omega$ dependence on time.
$\tau_h=3.0108~ \tau_{ba}=0.301,~\tau_c=3.014 ~~ \tau_{ch}=0.346   $.
The external parameters are: $\omega_c=5.382 ,~\omega_h=12.717,
~J=2.,~T_h=7.5,~T_c=1.5,~\Gamma_h=0.382,~\Gamma_c=0.342,
~\gamma_h=\gamma_c=0$
}
\label{fig:cycle1}
\end{figure}

The dynamics of the working medium has been described 
in Sec. \ref{sec:model}.
The parameters defining the cycle are:
\begin{itemize}
\item{$T_h$ and $T_b$, the hot/cold  bath temperatures.}
\item{ $\Gamma_h$ and $\Gamma_c$, the hot/cold bath heat conductance parameters.}
\item{$\gamma_h$ and $\gamma_c$, the hot/cold bath dephasing parameters.}
\item{$J$-the strength of the internal coupling}
\end{itemize}
The external control parameter define the four strokes of the cycle 
(Cf. Fig. \ref{fig:cycle1}):
\begin{enumerate}
\item{ {\em Isochore} $A \rightarrow B$: when the field   
is maintained  constant, $\omega=\omega_b$, the working medium
is in contact with the hot bath for a period of $\tau_h$. }
\item{ {\em Adiabat} $B \rightarrow C$: when the field changes linearly 
from $\omega_b$ to $\omega_a$ in a time period of $\tau_{ba}$.}
\item{ {\em Isochore} $C \rightarrow D$: when the field   is maintained  
constant $\omega=\omega_a$ the working medium
is in contact with the cold bath for a period of $\tau_c$. }
\item{ {Adiabat} $C \rightarrow A$: when the field changes linearly 
from $\omega_a$ to $\omega_b$ in a time period  of $\tau_{ab}$.}
\end{enumerate}
The trajectory of the cycle in the field  and the entropy plane ($\omega,{\cal S}_E$)
is shown in Fig. \ref{fig:cycle1} employing a numerical propagation 
with a linear $\omega$ dependence on time. 

\begin{figure}[tb]
\vspace{0.66cm}
\hspace{3.cm}
\psfig{figure=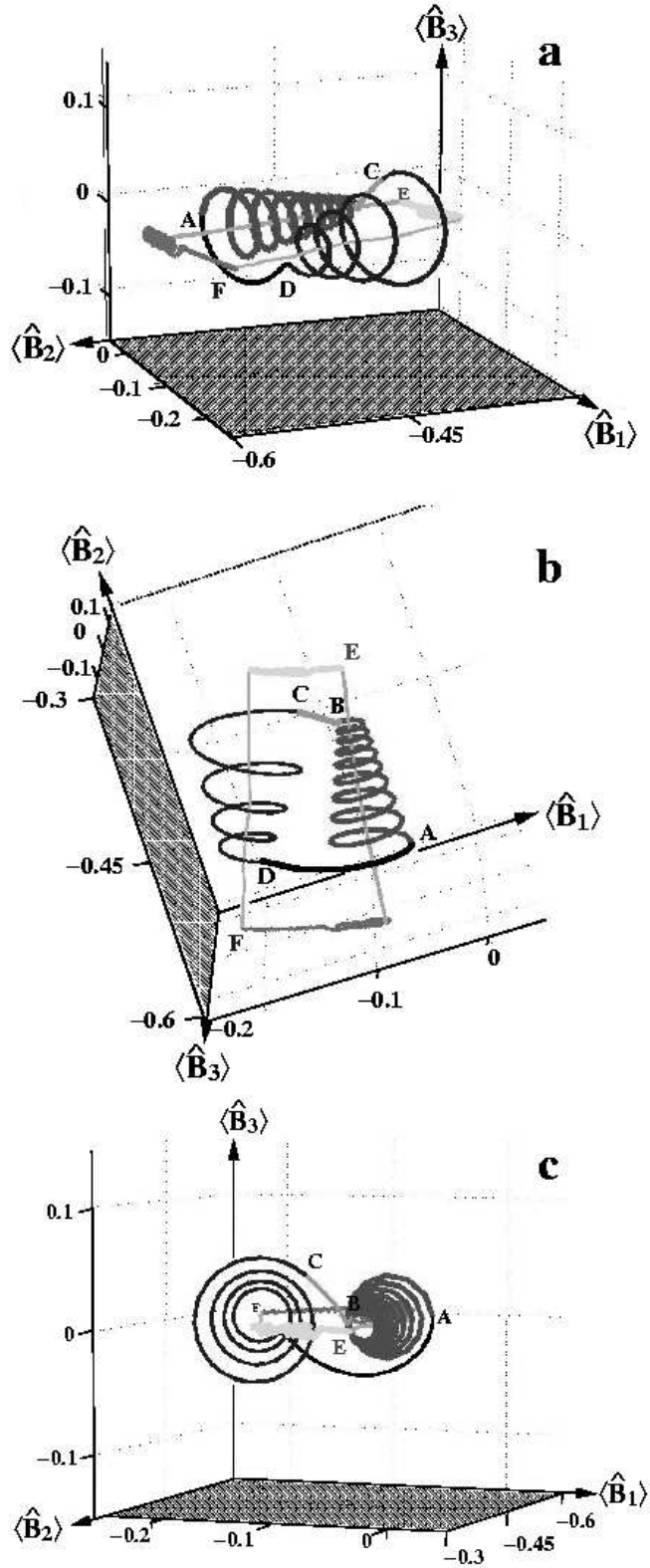,width=0.55\textwidth}
\vspace{0.5cm}
\caption{
The optimal cycle trajectory ABCD and the infinitely long trajectory EF
in the $b_1=\langle {\bf \hat B_1} \rangle ~,~b_2=\langle {\bf \hat B_2} \rangle ~,~
b_3=\langle {\bf \hat B_3} 
\rangle$
coordinate system showing three view points.}
\label{fig:cyctraj}
\end{figure}

A different perspective on the dynamics during the cycle of operation
is shown in Fig.  \ref{fig:cyctraj}, displaying  the cycle trajectory in the 
$b_1,b_2,b_3$ coordinates. 
The hypothetical cycle with infinitely long time on all branches
would include the equilibrium points E and F. The cycle trajectory is planar
on the ${\bf \hat B_3}=0 $ plane as can be seen in panel C. The 
cycle ABCD with finite time allocation spirals around the infinitely long time
cycle with an incursion into the  ${\bf \hat B_3}$ directions.
The reference cycle with infinite time allocation on all branches is characterized
by a diagonal state ${{{\boldsymbol{ \mathrm {  \rho}}}}}_e$ in the instantaneous energy representation.
The slow motion on the {\em adiabats} 
allows the state ${{{\boldsymbol{ \mathrm {  \rho}}}}}$  to adopt to the changes in time of the Hamiltonian,
which  therefore can be termed adiabatic following. If the time allocation
on the adiabats is short, non-adiabatic effects take place. In the sudden limit
of infinite short time allocation on the {\em adiabat}, the state of the system
has no time to evolve ${{{\boldsymbol{ \mathrm {  \rho}}}}}(t_i+\tau_{ab})={{{\boldsymbol{ \mathrm {  \rho}}}}}(t_i)$. 
The Hamiltonian will then change
from ${\bf \hat H_i} = \omega(t_i) {\bf \hat B_1}+J {\bf \hat B_2}$ to 
${\bf \hat H_f} = \omega(t_i+\tau_{ab}) {\bf \hat B_1}+J {\bf \hat B_2}$ therefore the representation
of the state ${{{\boldsymbol{ \mathrm {  \rho}}}}}_e (t_i+\tau_{ab})$ in the new energy representation is
rotated by an angle $\theta=(\theta_i-\theta_f)$ compared to the former one. Where,
$\theta_i=\arcsin(J/\Omega(t_i))$ and $\theta_f=\arcsin\left(J/\Omega(t_i+\tau_{ab})\right)$.
\begin{figure}[tb]
\vspace{.0cm}
\hspace{3.cm}
\psfig{figure=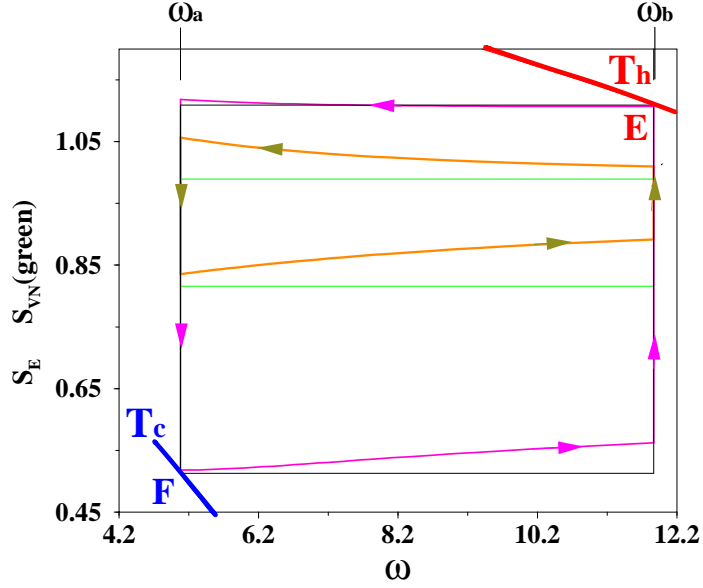,width=0.6\textwidth}
\vspace{0.5cm}
\caption{Three cycles of operation based on the analytic solution in the
($\omega, {\cal S}_E$) plane. The orange inner cycle has the shortest 
time allocations ($\tau_h=2.~ \tau_{ba}=\tau_{ab}=0.05, ~~ \tau_c=2.1   $).
The green cycle shows the corresponding $(\omega, {\cal S}_{VN})$ plot.
The magenta cycle has longer time allocations 
$\tau_h=\tau_c=15.~ \tau_{ba}=\tau_{ab}=0.015$, while the black
cycle has infinite time allocations on all branches therefore ${\cal S}_E~=~{\cal S}_{VN}$. 
This cycle touches the isothermal equilibrium points E and F.
The common parameters for all the cycles are:
$~J=2.,r=0.96,$$~T_h=7.5,~T_c=1.5,$$~\Gamma_h= \Gamma_c=0.3243,$$
 ~\gamma_h=\gamma_c=0,$$ \omega_a=5.08364,$$~\omega_b=11.8675$.
}
\label{fig:analytic} 
\end{figure}
When following the direction of the cycle, the energy-entropy increases on the {\em adiabts}.
This is evident in both Fig. \ref{fig:cycle1} and Fig. \ref{fig:analytic}. 
This entropy increase is the signature of nonadiabatic effects reflecting the inability
of the population on the energy states  to follow the change in time
of the Hamiltonain. As a result the energy dispersion increases.
Since the evolution on these branches is unitary, ${\cal S}_{VN}$ is constant.
When more time is allocated to the {\em adiabats} the increase in ${\cal S}_{E}$
is smaller. For infinite time allocation
${\cal S}_{E}~~=~~{\cal S}_{VN}$. In this case the state of the woking medium 
is always diagonal in the energy representation. The larger curvature of the entropy increase in
the analytic result of Fig. \ref{fig:analytic}, compared with
the numerical result of Fig. \ref{fig:cycle1} reflects the difference
in the dependence of $\omega(t)$ on time. When the analytic functional
form of $\omega (t)$ is used in the numerical propagation  the numerical
solution converges to the values of the analytic solution. This convergence test
was used as a consistancy check for both methods. Convergence was not uniform for
all elements in the propagator  (Cf. Eq. (\ref{propag}) and Eq. (\ref{propan}) ). 
Comparing the elements of the numerical propagator ${\cal U}_a(\tau_{ab})$ to 
the elements of analytic  ${\cal U}_a(\tau_{ab})$, showed that the largest discrepency between the
individual elements at $t=\tau_{ab}$ was less than $10^{-3}$ when 
a time step of $\Delta t= \tau_{ab}/1000$ was used.

\begin{figure}[tb]
\vspace{0.66cm}
\hspace{3.cm}
\psfig{figure=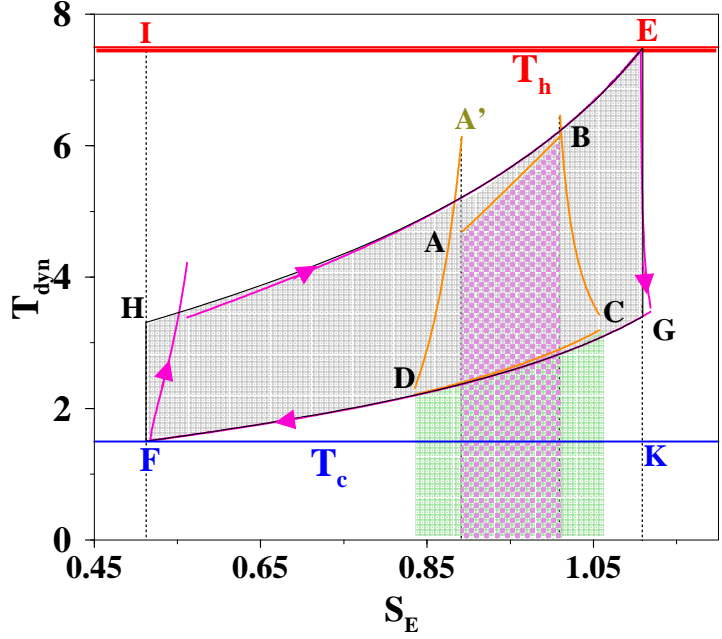,width=0.6\textwidth}
\vspace{0.5cm}
\caption{The cycles in
$({\cal S}_E,T_{dyn})$ planes.  The inner cycle A,B,C,D corresponds to the
short time cycle of Fig. (\ref{fig:analytic}). The magneta cycle is the long
time cycle and the black cycle H,E,G,F corrosponds to the cycle with infinite time 
allocation on all branches.
The rectangle, including points I,E,K,F is the work obtained in a Carnot
cycle operating between $T_h$ and $T_c$. The shaded area H,E,G,F represents the maximum
work of the Otto cycle. The area below the A,B segment is the heat 
trasfered from the hot bath ${\cal Q}_h$. The area below the D,C segment is the heat
transfered to the cold bath ${\cal Q}_c$.}
\label{fig:3intemp}
\end{figure}

In Fig. \ref{fig:3intemp} the cycle of operation is presented 
in the energy-entropy internal-temperature
coordinates $({\cal S}_E,T_{dyn})$. 
The cycles shown  corresponds to the analyticial cycles of Fig. \ref{fig:analytic}.
The discontinuities in the short time cycle reflect over-heating in the compression stage
as shown as the difference between the point A and A' in Fig. \ref{fig:3intemp}.
The heat accumulated is quenched when the working medium is put in contact with the hot bath.
This phenomena has been identified in measurements of working fluid temperatures
in actual heat engines or heat pumps \cite{gordon98}. A discontinuity
as a result of insufficient cooling of the woking medium
in the expansion branch is also evident in the short time cycle.
The magnitude of these discontinuities is reduced at longer times
and dissapear for the infinite long cycle where the working fluid
reaches thermal equilibrium with the hot bath at point $\bf E$ and with the cold
bath at point $\bf F$. In this case both {\em adiabatic} branches are isoentropic.
It is clear from Fig. \ref{fig:3intemp}, that for the cycles with vertical {\em adiabats}
the work is the area enclosed by the cycle trajectory. When the time allocation
on the {\em adiabats} is restricted this is no longer the case since due to the entropy
increase, the area under the hot {\em isochore} does not cover the area under the cold 
{\em isochore}. Additional cooling is then required to dissipate the extra work 
required to drive the system on the {\em adiabats} at finite time.

\section{The effect of phase and dephasing.}
\label{sec:phase}

The performence of the heat engine explicitly depend on 
heat and work which constitute the energy (\ref{eq:firstlaw}). Do other observables,
incompatiable with the energy, influence the engins performence?
Examining the cycle trajectory on the {\em isochores} in Fig. \ref{fig:cyctraj}, 
in addition to the motion in the energy direction, towoard 
equilibration, spiraling motion exists.
This motion is characterized by amplitude and phase of an observable in the plane 
perpendicular to the energy direction.  The phase $\phi$ of this motion advances in time, 
i.e. $\phi \propto t$.
The concept of phase has its origins in classical mechanics where a canonincal
trnsformation leads to a new set of action angle  variables. 
The conjugate variable to the Hamiltonain is the phase.
In quantum mechanics the phase observable has been a subject of contineous debate
\cite{nieto68}. For a harmonic oscillator it is related to the creation and anhilation
operator $\bf \hat a$ \cite{levyleblond,gour02}. In analogy  the
raising/lowering operator is defined:
\begin{equation}
{\bf \hat L}_{\pm} ~~=~~\frac{1}{\sqrt{2}\Omega}\left( -J {\bf \hat B_1} + \omega {\bf \hat B_2} 
\pm i \Omega {\bf \hat B_3}\right)~~~,
\label{eq:Lpm}
\end{equation}
which has the following commutation relation with the Hamiltonian:
\begin{equation}
[{\bf \hat H},{\bf \hat L}_{\pm}] = \pm  \sqrt{2} \Omega {\bf \hat L}_{\pm}~~~.
\label{eq:lham}
\end{equation}
The free evolution of $\bf \hat L_{+}$  therefore becomes: 
$\bf \hat L_{+}(t) = e^{i \sqrt{2}\Omega t}{\bf \hat L}_{+}(0)$ which defines the phase
variable through: $ \langle {\bf \hat L_+} \rangle ~=~ r e^{i \phi}$,
therefore $\phi= \arctan \left( \frac{\Omega b_3}{-J b_1+\omega b_2} \right)$.
A corroboration for this interpretation is found by examining the state ${{{\boldsymbol{ \mathrm {  \rho}}}}}_e$
in the energy representation (Cf. Eq. (\ref{rorop}). 
The off diagonal elements are completely specified by the expectation values of ${\bf \hat L}_{\pm}$.

The dynamics  of ${\bf \hat L}_{\pm}$ on the  {\em isochores} includes also 
dissipative contributions which can be evaluated using Eq. (\ref{expmowithg}):
\begin{equation}
\dot {\bf \hat L}_{\pm}~~=~~ \pm i \sqrt{2} \Omega {\bf \hat L}_{\pm} 
-\left( \Gamma + 2 \gamma \Omega^2 \right) {\bf \hat L}_{\pm}
\label{eq:ldynm}
\end{equation}
Examining Eq. (\ref{eq:ldynm}) it is clear that the amplitude of $ {\bf \hat L}_{\pm}$
decays exponentially with the rate $\frac{1}{T_2}=\Gamma +  2 \gamma \Omega^2$, where
$\Gamma$ is the dephasing contribution due to energy relaxation and  
$\frac{1}{T_2^*}=2 \gamma \Omega^2$ is the pure dephasing contribution.
\begin{figure}[tb]
\vspace{0.5cm}
\hspace{3.cm}
\psfig{figure=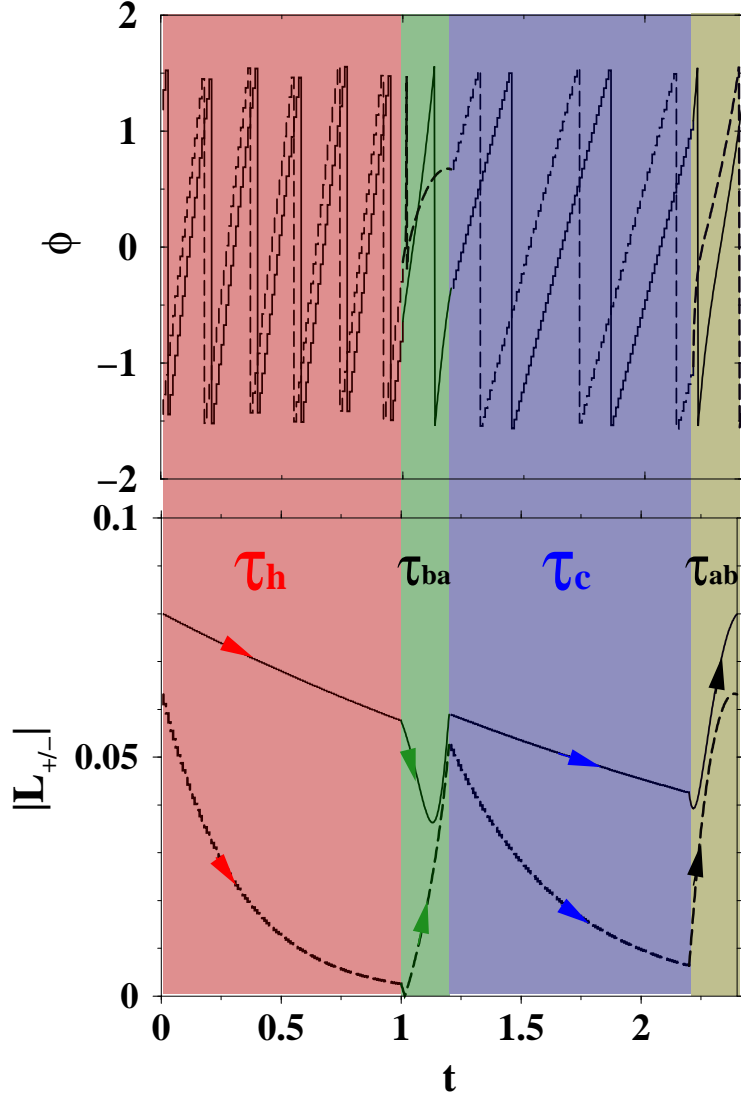,width=0.6\textwidth}
\caption{The modulus and phase of ${\bf \hat L}_{\pm}$ as a function of time.
The dashed lines include additional pure dephasing ($\gamma_h=0.01,\gamma_c=0.03$).
The common parameters are: $T_h$=7.5, $T_c=1.5$,$\Gamma_h=\Gamma_c=0.34$,
$\omega_b=11.8675, \omega_a=5.083$,
The total cycle time is $\tau=2.4$ where,
$ \tau_{h}=\tau_{c}=1$, $ \tau_{ba}=0.2, \tau_{ab}=0.2$.}
\label{fig:phase6}
\end{figure}

\begin{figure}[tb]
\vspace{0.5cm}
\hspace{3.cm}
\psfig{figure=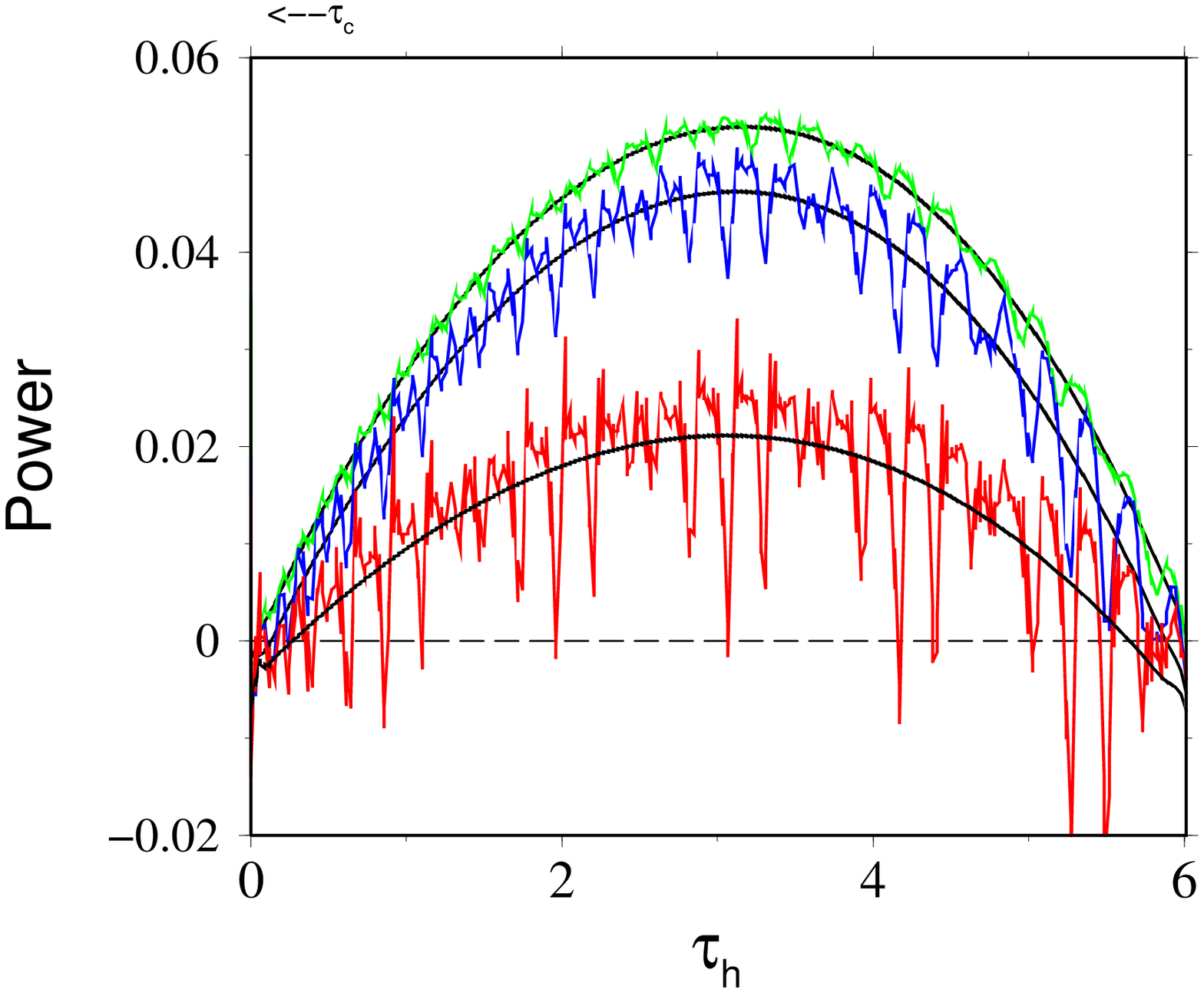,width=0.6\textwidth}
\caption{The power produced by the engine as a function of the time allocation on the hot {\em isochore}.
For the green cycle J=1 and $\Gamma_h = \Gamma_c=0.324$.
For the blue cycle J=2 and $\Gamma_h = \Gamma_c=0.324$.
For the red cycle  J=2 and $\Gamma_h = \Gamma_c=0.162$.
The three colored cycles have no pure dephasing $ \gamma_h=\gamma_c= 0 $.
With addition of dephasing $ \gamma_h=0.01 $ and  $ \gamma_c=0.03 $ 
the "noise" is eliminated and the three cycles collapse to the solid
black lines.
The common parameters are: $T_h$=7.5, $T_c=1.5$,
$\omega_b=12.717, \omega_a=5.382$,
The total cycle time $\tau$ is:=6.74,
$ \tau_{ba}=0.3, \tau_{ab}=0.34$.}
\label{fig:pwoer6}
\end{figure}

\begin{figure}[tb]
\vspace{0.5cm}
\hspace{3.cm}
\psfig{figure=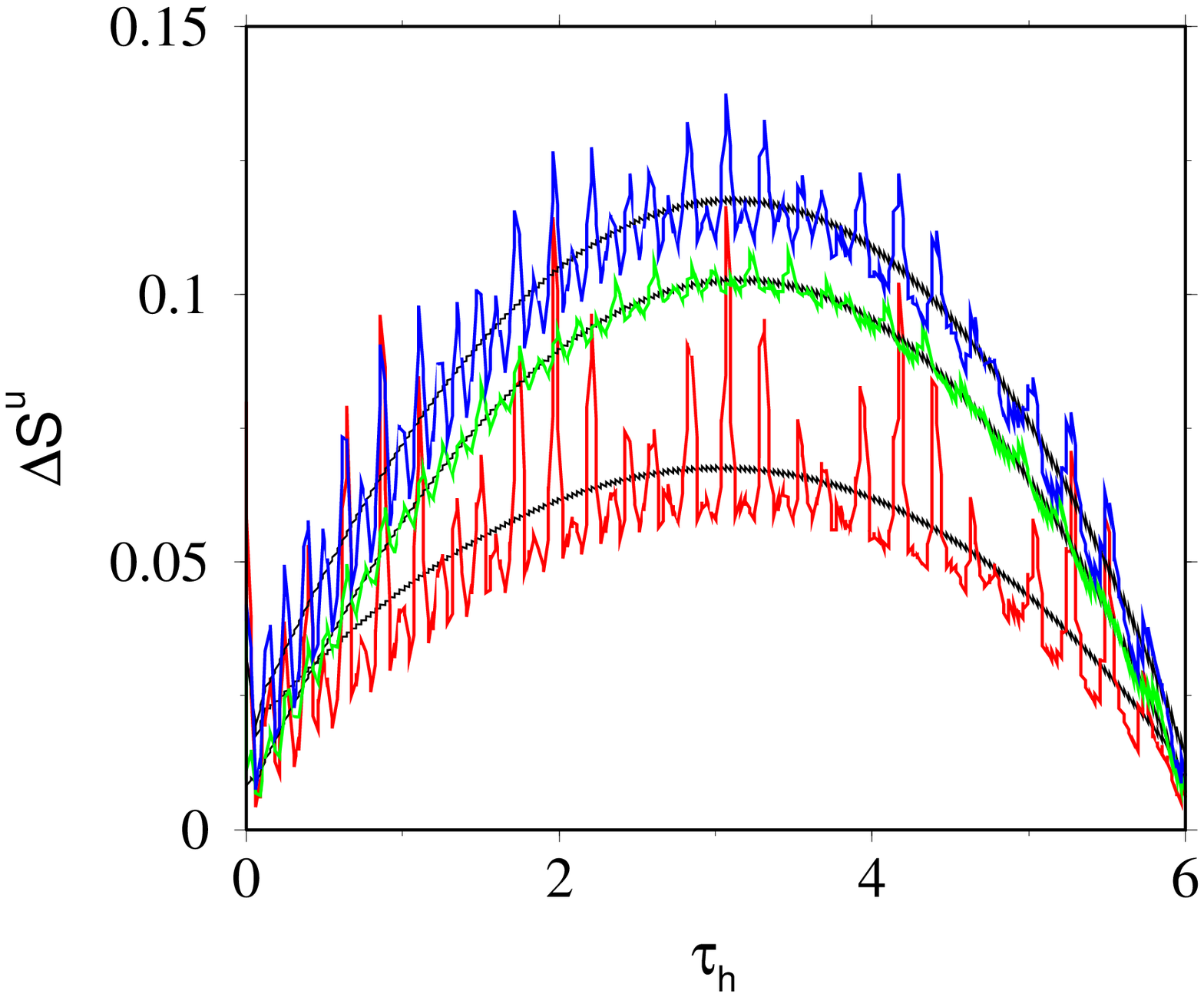,width=0.6\textwidth}
\vspace{0.cm}
\caption{Entropy productions ${\cal D S}_{cycle}$ Eq. (\ref{entpro}), as a function of the time
allocation on the  hot {\em isochore}. 
The notations are the same as Fig. \ref{fig:pwoer6}. }
\label{fig:entprod6}
\end{figure}

\begin{figure}[tb]
\vspace{1.0cm}
\hspace{3.cm}
\psfig{figure=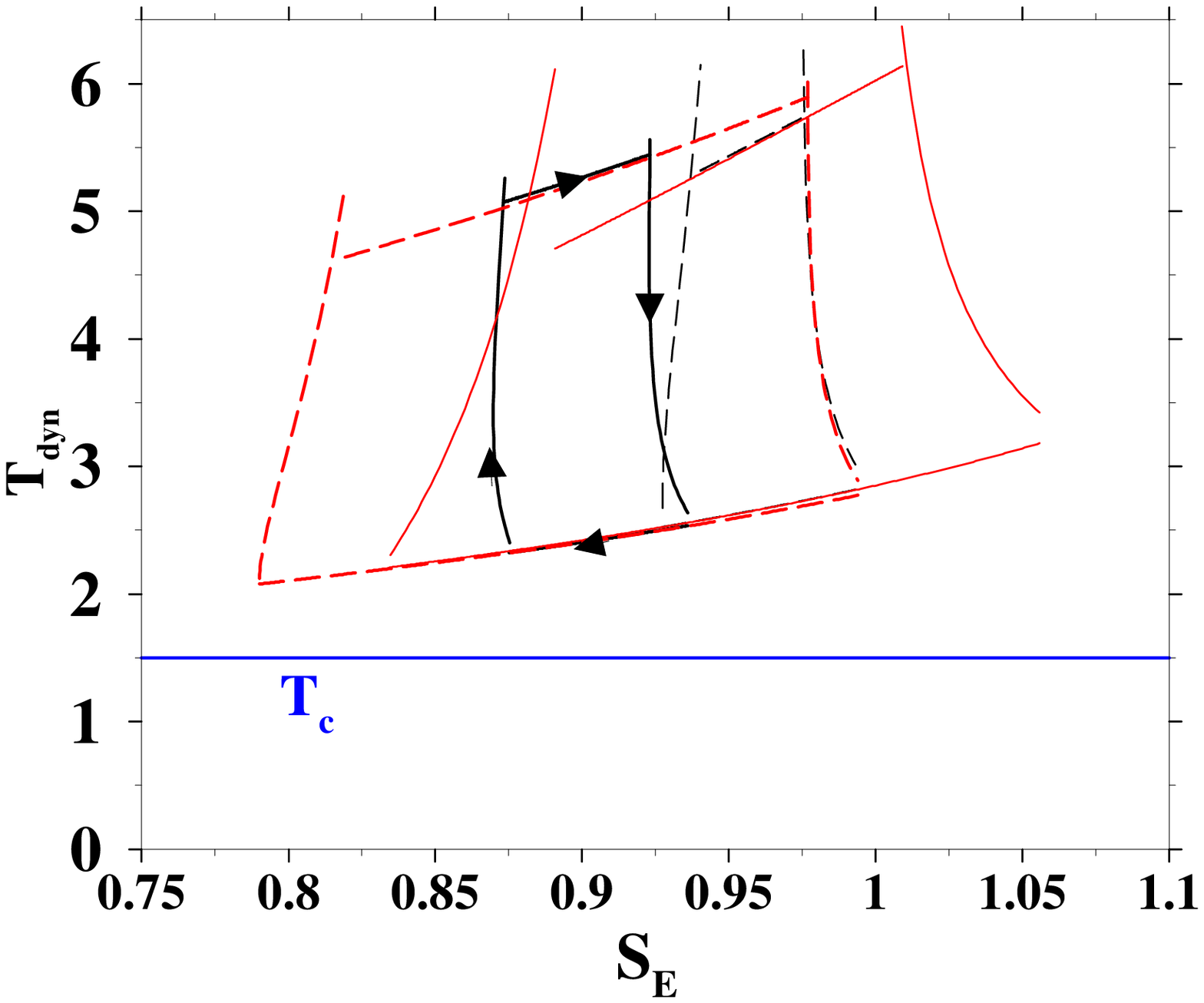,width=0.6\textwidth}
\vspace{0.5cm}
\caption{The influence of dephasing on the cycle of operation
in the $( {\cal S}_E ,  T_{dyn} )$ plane.
Solid curves correspond to an operation without pure dephasing,
The dashed curves represent cycles including pure dephasing.
For the black cycles the time allocations on the
{\em isochores} are: $ \tau_h=\tau_c=0.6$. The pure dephasing parameter is
$\gamma_h=\gamma_c=0$ for the solid lines, and 
$\gamma_h=0.005, \gamma_c=0.015$ for the dashed lines. 
For the red cycles the allocated times on the
isochores are: $ \tau_h=2., \tau_c=2.1$ with 
$\gamma_h=\gamma_c=0$ for the solid lines, and 
 $\gamma_h=0.01, \gamma_c=0.03$ for the dashed lines.
The common parameters for all four cycles are:
$J=2.$,$ T_h=7.5$,$ T_c=1.5$,$ \Gamma_h=\Gamma_c=0.3243$,
$\tau_{ab}=\tau_{ba}=0.015$.} 
\label{fig:4intemdeph}
\end{figure}

Both Fig. \ref{fig:cyctraj} and Fig. \ref{fig:phase6} show that the dephasing
is not complete at the end of the {\em isochores}. A small change in the time
allocation  in the order of $1 /\Omega$  
can completely change the final phase on the {\em isochore} and on the initial phase
for the {\em adiabat}. This means that the cycle performance characteristic 
becomes very sensitive to small changes in time allocation on the {\em isochores}.
This effect can be observed in Fig. \ref{fig:pwoer6} for 
the power and Fig. \ref{fig:entprod6} for the entropy production.
Examining Fig. \ref{fig:pwoer6} reveals that increasing $J$ increases the "phase" effect.
For $J=2$ for specific time allocations the power can even become negative.
Increasing the dephasing rate either by adding pure dephasing or
by changing the heat transfer rate reduces the "noise". This can also be seen in Fig. 
\ref{fig:entprod6}.
An interesting phase effect can be observed in Fig. \ref{fig:4intemdeph} 
where the cycle is displayed in the $({\cal S}_E,T_{dyn})$ plane.
The inner (solid black) cycle shows an energy-entropy decrease in the 
compression {\em adiabat}. The reason for this decrease is a phase memory
from the compression {\em adiabat} which is due to insufficient dephasing on the cold
{\em isochore}. Additional pure dephasing eliminates this entropy decrease
as can be seen in the dashed black cycle. 
This cycle is also pushed to larger entropy values. 
The orange cycles are characterized by a longer time allocation
on the {\em isochores}. For these cycles the energy-entropy 
always increases on the {\em adiabats}. This cycle is shifted
by dephasing to lower energy-entropy values. 

\section{discussion}
\label{sec:discuss}

Quantum thermodynamics is the study of thermodynamical phenomena 
based on quantum mechanical principles \cite{hoffmann}. To meet this challenge, 
quantum expectation values have to be related to thermodynamical variables.
The Otto cycle is an ab-initio quantum model for which analytic solutions 
have been  obtained. The principle thermodynamical variables: energy entropy and temperature
are derived from first principles. The solution of the quantum equations of motion
for the state ${{{\boldsymbol{ \mathrm {  \rho}}}}}$, enables tracing the thermodynamical
variables for each point on the cycle trajectory. This dynamical picture supplies a rigorous
formalism for {\bf finite-time-thermodynamics} \cite{salamon77,andresen83}.

An underlying principle  of finite-time-thermodynamics is that 
operation irreversibilities are inevitable if a process is run at finite rate.
Moreover these irreversibilities are the source of performance limitations 
imposed on the process.
The present Otto cycle heat engine in line with FTT is subject to two 
major performance limitations:
\begin{itemize}
\item{Finite rate of heat transfer from the hot bath to the working medium
and from the working medium to the cold bath.}
\item{Additional work invested in the expansion and compression branches
is required to drive the {\em adiabats} at a finite time.}
\end{itemize}
The finite rate of heat transfer limits the maximum obtainable power ${\cal P}$
\cite{curzon75}. The present Otto engine model
is not an exception, showing similarities with
previous studies of discrete quantum heat engines 
\cite{feldmann96,feldmann00,geva0,geva1}. 

The irreversibility caused by the finite time duration on the {\em adiabats}
is the novel finding of the present study
as well as  the preceding short letter \cite{kosloff01}.
This irreversibility is closely linked to the quantum adiabatic condition.
The nonadiabatic irreversibility is caused by the interplay of
the noncommutability of the Hamiltonian at different points
along the cycle trajectory and the dephasing caused by coupling 
to the heat baths on the {\em isochores}. In the present Otto cycle these 
contributions are separated in time. The non-adiabaticity can be characterized
by an increase in the modulus of $\langle {\bf \hat L}_{\pm} \rangle $ on 
the {\em adiabats}. Dephasing, i.e. exponential decay of the modulus of
$\langle {\bf \hat L}_{\pm} \rangle $ is induced by the coupling to the baths 
on the {\em isochores}.

The dynamics of the ${\bf \hat L}_{\pm}$ operator associated with the phase
can be compared to the ${\bf \hat B}_{\pm}$ operator associated with the internal correlation
between the spins (Cf. (\ref{rorop1})). The absolute value of $| \bf \hat B_{\pm} |$ oscillates 
on all branches of the cycle never reaching zero. This is not surprising since $\bf \hat B_{\pm}$ does not commute
with the Hamiltonian. The "angle" $\phi_{B}=\arctan({b_3/b_2})$ is excited for small cycle times. 
For cycles with large time allocation on the {\em isochores}, $\phi_{B}$ is found to be close to zero.
These observations reflect the two types of correlations between particles.
A "classical" correlation and a quantum correlation
meaning EPR \cite{EPR,lindblad73} entanglement between particles.
The general trend is therefore for the engine to become more "classical" when the
cycle times become longer. In this case the state follows the energy direction
and in addition entanglement between particles is small. Adding pure dephasing
has a similar effect. A continuous  measurement of energy during operation will also
lead to effective pure dephasing. For short cycle times quantum effects become important.
The entropy decrease on the adiabats which is the result of "phase" memory is such an example.
The quantum effect which influences the performance is the excess work on the {\em adiabat} due to
the inability of the state to follow the energy direction.

\acknowledgments

This research was supported by  the  US Navy under contract 
number N00014-91-J-1498 and the Israel Science Foundation.
The authors wish to thank Lajos Diosi and Jeff Gordon for their continuous support
and help.

\appendix

\section{The F operators}
\label{sec:foperators}

The method of construction of $\bf \hat F_{\rm j}$ is based on identifying the operators 
with the raising and lowering operators in the energy frame.
The matrix ${\cal C}$ which diagonalizes
the Hamiltonian becomes:
\begin{eqnarray}
\begin{array}{c}
{\cal C}
\end{array}~~=~\left( 
\begin{array}{cccc}
-\sqrt{\frac{\Omega - \omega}{ 2 \Omega}}  &0 & 0 & 
\sqrt{\frac{\Omega + \omega}{ 2 \Omega}}  \\
 0 &1 & 0 & 0 \\
 0 &0 & 1 & 0  \\
\sqrt{\frac{\Omega + \omega }{ 2 \Omega}}  & 0 & 0 & 
\sqrt{\frac{\Omega - \omega }{ 2 \Omega}}  \\
\end{array} 
\right)
\label{EIGHA}
\end{eqnarray}
Denoting $\sqrt{\frac{\Omega - \omega }{ 2 \Omega}}~=~\mu~~$, and
$\sqrt{\frac{\Omega + \omega }{ 2 \Omega}}~=~\chi~~~$, the diagonalization 
of the Hamiltonian matrix becomes:
\begin{eqnarray}
\left(
\begin{array}{cccc}
-\mu&0 & 0&\chi \\
0&1& 0&0 \\
0 &0&1&0 \\
\chi&0 & 0&\mu \\
\end{array} 
\right)
\left(
\begin{array}{cccc}
\frac{\omega }{ \sqrt{2}} &0&0&\frac{J }{ \sqrt{2}} \\
0&0&0 &0\\
0&0&0 &0\\
\frac{J }{ \sqrt{2}}   &0 & 0& -\frac{ \omega }{ \sqrt{2}}  \\
\end{array} 
\right)
\left(\begin{array}{cccc}
-\mu&0 & 0&\chi \\
0&1& 0&0 \\
0 &0&1&0 \\
\chi&0 & 0&\mu \\
\end{array} 
\right)~~=~~\left(
\begin{array}{cccc}
-\frac{ \Omega }{ \sqrt{2} }&0&0&0 \\
0&0&0&0 \\
0&0&0&0 \\
0&0&0&\frac{ \Omega }{ \sqrt{2}} \\
\end{array} \right)
\label{DIAGham}
\end{eqnarray}
The down transition rates $k \downarrow$ are chosen to be equal for all the four
transitions, while the  raising transitions  $k \uparrow$ 
comply with detailed balance. Schematically the eight 
transitions are:
\begin{eqnarray}
\begin{array}{cccccccc}
\bf F_1 &\bf F_2 &\bf F_3&\bf F_4&\bf F_5&\bf F_6&\bf F_7&\bf F_8 \rm\\
E_2 &E_2 &E_3&E_3&E_4&E_4&E_4&E_4 \\
\Uparrow & \Downarrow & \Uparrow& \Downarrow& \Uparrow& \Downarrow&
\Uparrow& \Downarrow \\
E_1 &E_1&E_1&E_1&E_2&E_2&E_3&E_3 \\
\end{array} 
\label{RANS1}
\end{eqnarray}

\subsubsection{Detailed presentation of a few $F_i$ operators.}

The $\bf \hat F$ operator for the transition $E_1$ to $E_2$, is
$\bf F_{1 \rightarrow 2} \equiv \bf F_1$;  
In the energy picture, it is simply:
\begin{eqnarray}
\begin{array}{c}
\bf F_1~~=~~
\end{array}
\sqrt{k \downarrow}  \left(\begin{array}{cccc}
0&0 & 0&0 \\
1&0& 0&0 \\
0 &0&0&0 \\
0&0 &0&0 \\
\end{array} 
\right)
\label{repF1}
\end{eqnarray}
Using the matrix ${\cal C}$  to transform back to the
polarization picture leads to:
\begin{eqnarray}
\nonumber
\begin{array}{c}
\bf F_1~~=~~
\end{array}
\sqrt{k \downarrow}  \left(\begin{array}{cccc}
-\mu &0 & 0& \chi \\
0&1& 0&0 \\
0 &0&1&0 \\
\chi &0 & 0& \mu \\
\end{array} 
\right)
\left(\begin{array}{cccc}
0 &0&0&0 \\
1&0&0 &0\\
0&0&0 &0\\
0&0 & 0&0 \\
\end{array} 
\right)
\left(\begin{array}{cccc}
-\mu &0 & 0& \chi \\
0&1& 0&0 \\
0 &0&1&0 \\
\chi &0 & 0& \mu \\
\end{array} 
\right)~~=~
\end{eqnarray}
\begin{eqnarray}
\sqrt{k \downarrow}    \left(
\begin{array}{cccc}
0&0&0&0 \\
-\mu &0&0& \chi \\
0&0&0&0 \\
0&0&0&0 \\
\end{array} \right)
\label{TrF1}
\end{eqnarray}
And $\bf F_1^{\dagger}$ will be;
\begin{eqnarray}
\nonumber
\begin{array}{c}
\bf F_1^{\dagger}~~=~~
\end{array}
\sqrt{k \downarrow}  \left(\begin{array}{cccc}
-\mu &0 & 0& \chi \\
0&1& 0&0 \\
0 &0&1&0 \\
 \chi &0 & 0& \mu \\
\end{array} 
\right)
\left(\begin{array}{cccc}
0 &1&0&0 \\
0&0&0 &0\\
0&0&0 &0\\
0&0 & 0&0 \\
\end{array} 
\right)
\left(\begin{array}{cccc}
-\mu &0 & 0& \chi \\
0&1& 0&0 \\
0 &0&1&0 \\
\chi &0 & 0& \mu \\
\end{array} 
\right)~~=~~
\end{eqnarray}
\begin{eqnarray}
\sqrt{k \downarrow}    \left(
\begin{array}{cccc}
0&-\mu &0&0 \\
0&0&0&0 \\
0&0&0&0 \\
0& \chi &0&0 \\
\end{array} \right)
\label{TF1D}
\end{eqnarray}
Using a similar procedure all the $\bf \hat F_i$ in the polarization picture become: 
\begin{eqnarray}
\begin{array}{c}
{\bf \hat F_1}~~=~~{\bf \hat F}_{1 \rightarrow 2}~~=~~
\end{array}
\sqrt{k \downarrow}  \left(\begin{array}{cccc}
0&0 & 0&0 \\
-\mu &0& 0& \chi \\
0 &0&0&0 \\
0&0 &0&0 \\
\end{array} 
\right)
\label{repFF1}
\end{eqnarray}

\begin{eqnarray}
\begin{array}{c}
{\bf \hat F_2}~~=~~{\bf \hat F}_{2 \rightarrow 1}~~=~~
\end{array}
\sqrt{k \uparrow}  \left(\begin{array}{cccc}
0&-\mu & 0&0 \\
0&0& 0&0 \\
0 &0&0&0 \\
0& \chi &0&0 \\
\end{array} 
\right)
\label{repFF2}
\end{eqnarray}

\begin{eqnarray}
\begin{array}{c}
{\bf \hat F}_3~~=~~\bf F_{1 \rightarrow 3}~~=~~
\end{array}
\sqrt{k \downarrow}  \left(\begin{array}{cccc}
0&0 & 0&0 \\
0&0& 0&0 \\
-\mu &0&0& \chi \\
0&0 &0&0 \\
\end{array} 
\right)
\label{repFF3}
\end{eqnarray}

\begin{eqnarray}
\begin{array}{c}
{\bf \hat F_4}~~=~~{\bf \hat F}_{3 \rightarrow 1}~~=~~
\end{array}
\sqrt{k \uparrow}  \left(\begin{array}{cccc}
0&0 &-\mu &0 \\
0&0& 0&0 \\
0 &0&0&0 \\
0&0 & \chi &0 \\
\end{array} 
\right)
\label{repFF4}
\end{eqnarray}

\begin{eqnarray}
\begin{array}{c}
\Op F_5~~=~~\Op F_{2 \rightarrow 4}~~=~~
\end{array}
\sqrt{k \downarrow}  \left(\begin{array}{cccc}
0& \chi & 0&0 \\
0&0& 0&0 \\
0 &0&0&0 \\
0& \mu &0&0 \\
\end{array} 
\right)
\label{repFF5}
\end{eqnarray}

\begin{eqnarray}
\begin{array}{c}
\Op F_6~~=~~\Op F_{4 \rightarrow 2}~~=~~
\end{array}
\sqrt{k \uparrow}  \left(\begin{array}{cccc}
0&0 &0&0 \\
\chi &0& 0& \mu \\
0 &0&0&0 \\
0&0 &0&0 \\
\end{array} 
\right)
\label{repFF6}
\end{eqnarray}

\begin{eqnarray}
\begin{array}{c}
\Op F_7~~=~~\Op F_{3 \rightarrow 4}~~=~~
\end{array}
\sqrt{k \downarrow}  \left(\begin{array}{cccc}
0&0 & \chi &0 \\
0&0& 0&0 \\
0 &0&0&0 \\
0&0 & \mu &0 \\
\end{array} 
\right)
\label{repFF7}
\end{eqnarray}

\begin{eqnarray}
\begin{array}{c}
\Op F_8~~=~~\Op F_{4 \rightarrow 3}~~=~~
\end{array}
\sqrt{k \uparrow}  \left(\begin{array}{cccc}
0&0 &0&0 \\
0&0& 0&0 \\
\chi &0&0& \mu \\
0&0 &0&0 \\
\end{array} 
\right)
\label{repFF8}
\end{eqnarray}

\end{document}